\documentclass[usenatbib,useAMS,usegraphicx,apj,iop]{emulateapj}
\usepackage{amssymb}
\usepackage{times}

\shorttitle{Foretellings of Ragnar\"ok}
\shortauthors{A. J. Mustill \& E. Villaver}
\begin{document}

\title{Foretellings of Ragnar\"ok: World-engulfing Asymptotic Giants and the Inheritance of White Dwarfs}

\author{Alexander J. Mustill, Eva Villaver}
\affil{Departamento de F\'isica Te\'orica, Universidad Aut\'onoma de Madrid, Cantoblanco, 28049 Madrid, Espa\~na}
\email{alex.mustill@uam.es}

\begin{abstract}
The search for planets around White Dwarf stars, and evidence for dynamical instability around them in the form of atmospheric pollution and circumstellar discs, raises questions about the nature of planetary systems that can survive the vicissitudes of the Asymptotic Giant Branch (AGB). We study the competing effects, on planets at several AU from the star, of strong tidal forces arising from the star's large convective envelope, and of the planets' orbital expansion due to stellar mass loss. We, for the first time, study the evolution of planets while following each thermal pulse on the AGB. For Jovian planets, tidal forces are strong, and can pull into the envelope planets initially at $\sim3$\,AU for a $1\mathrm{\,M}_\odot$ star and $\sim5$\,AU for a $5\mathrm{\,M}_\odot$ star. Lower-mass planets feel weaker tidal forces, and Terrestrial planets initially within $1.5-3$\,AU enter the stellar envelope. Thus, low-mass planets that begin inside the maximum stellar radius can survive, as their orbits expand due to mass loss. The inclusion of a moderate planetary eccentricity slightly strengthens the tidal forces experienced by Jovian planets. Eccentric Terrestrial planets are more at risk, since their eccentricity does not decay and their small pericentre takes them inside the stellar envelope. We also find the closest radii at which planets will be found around White Dwarfs, assuming that any planet entering the stellar envelope is destroyed. Planets are in that case unlikely to be found inside $\sim1.5$\,AU of a White Dwarf with a $1\mathrm{\,M}_\odot$ progenitor and $\sim10$\,AU of a White Dwarf with a $5\mathrm{\,M}_\odot$ progenitor.
\end{abstract}

\keywords{planets and satellites: dynamical evolution and stability --- stars: AGB and post-AGB --- stars: evolution}

\section{Introduction}

The discovery of White Dwarf stars with metal-polluted atmospheres or with circumstellar discs strongly suggests that planets, feeding asteroidal or cometary material to locations close to the star, exist around these degenerate stellar remnants \citep{Zuckerman+03,Gaensicke+06,Farihi+09,Gaensicke+12}. Furthermore, there are claims of planetary detection, by means of variations in the timing of stellar pulsations or binary eclipses, around various kinds of evolved star \citep[e.g.,][]{Mullally+08,Qian+09}. For such planets to remain to late times they must, however, first survive the late stages of stellar evolution. During the giant phases of the host star, the star loses a large fraction of its mass, causing the planet's orbit to expand in order that angular momentum be conserved \citep{Hadjidemetriou63,Alexander+76,LivioSoker84,VL07,VL09,Veras+11}. Countering this, however, the stellar envelope expands to a radius of several AU, and tidal forces wax stronger, potentially drawing planets in to their destruction in the envelope \citep{Rasio+96,VL07,VL09}. Dynamically, the survival of planets with semi-major axes of a few AU is governed by the balance between the tidal force which pulls the planet towards the expanded envelope, and the effects of stellar mass loss which push the planet away.

The fate of planets around Red Giant Branch (RGB) stars has received much attention, since many planets have been discovered around sub-giant, RGB and horizontal branch stars (e.g., \citealt{Niedzielski+09,Bowler+10,Johnson+11,Wittenmyer+11}; a recent compendium is given in \citealt{Gettel+12}). Such stars afford the only way to probe, with radial velocity surveys, the population of planets at radii of up to a few AU around stars more massive than the Sun, and a lack of close-in planets around these stars has been found. The cause of this deficiency has been attributed both to different planet formation processes around these stars \citep{Currie09}, and to tidal engulfment of close-in planets as the star expands and ascends the RGB \citep{Sato+08,VL09}. Regardless of the planetary population on entering the RGB, studies find that after the RGB phase planets around $1\mathrm{\,M}_\odot$ stars should have survived if they are located beyond a few AU, while around more massive stars the minimum radii at which planets should have survived are smaller, since the more massive stars have much smaller maximum radii on the RGB than Solar mass stars \citep{Sato+08,VL09,Kunitomo+11}. The survival radius increases with planet mass as tidal forces are stronger for more massive planets \citep{VL09}. There should, therefore, be a large population of planets at radii of several AU which will survive until the asymptotic giant branch (AGB) phase of evolution. A few planets, moreover, have been found in tight orbits around stars on the horizontal branch \citep{Silvotti+07,Charpinet+11,Setiawan+10}.  These planets could have survived the evolution of the star along the RGB, in the process possibly triggering mass loss to  form an sdB star \citep{Geier+09,Heber09,Soker98}, or could be the remnants of the tidally destroyed metallic cores of massive planets \citep{BearSoker12}. 

The population of planets around White Dwarf stars is, however, completely unknown. Despite the advantages that White Dwarfs offer to  direct detection of massive planets in the infrared \citep{Burleigh+02}, and the heroic efforts in the search for these planets using this technique \citep{Hogan+09,Steele+11} and others \citep{Mullally+08,Debes+11,Faedi+11}, no planet has so far been confidently detected orbiting a White Dwarf. Studies of planetary dynamics during the preceding AGB phase can therefore play a useful role by helping to inform such searches by predicting the properties of planets that survive the AGB.

On the AGB is attained, for many stars, the greatest mass loss; this, coupled with the large stellar radius and therefore strong tidal forces, means that the AGB phase is likely to be the key determinant in the survival of planets to the WD stage. The survival of planets during this phase has, however, received less attention. The survival of giant planets during and after the AGB was investigated by \cite{VL07}, in which study were included the effects of stellar mass-loss, the expanding stellar envelope, and evaporation of the planet during the Planetary Nebula phase. Tidal effects were, however, neglected. These were included in a more recent study \citep{Nordhaus+10}, which followed the evolution of giant planets and brown dwarfs from the ZAMS to the WD phase under tidal evolution and stellar mass loss using the \cite{Reimers75} prescription.

In the present paper we extend previous work on the fate of planets around AGB stars, following the planets' orbital evolution under the opposing effects of stellar mass loss and tidal forces. We make use of realistic stellar evolutionary models that follow the star's evolution through each thermal pulse, during which the stellar radius experiences large variations, and that have a mass loss prescription tested to be valid during this phase. This enables us to calculate the changing strength of the tides as the star evolves. We extend the previous studies to planets of lower mass, which can be dynamically significant on the WD stage in their interactions with remnant planetesimals \citep*{BMW11}. Also considered are the effects of including a moderate planetary eccentricity, and of changing the several parameters in the tidal formalism.

This paper is organized as follows. In Section~\ref{sec:eqns} we set out the equations governing the evolution of the planet's orbit. In Section~\ref{sec:models} we describe the stellar models used. The results of our numerical integrations are presented in Section~\ref{sec:numerics}, and discussed in Section~\ref{sec:discussion}. We conclude in Section~\ref{sec:conclusions}.

\section{Governing equations}

\label{sec:eqns}

We adopt the formalism of \cite{Zahn77} for the tidal forces raised on the star by the planet. The dominant tidal effect is the damping of the equilibrium tide by viscous dissipation in the star's large convective envelope, which spans almost the entire stellar radius (see Section~\ref{sec:models}). It has been shown \citep{VerbuntPhinney95} that this is a good model of tidal forces in stellar binaries with a giant component, but it is unclear whether it will be accurate for the weaker tides raised by planets. We also consider tidal dissipation in the planet, but shall find that it is unimportant. We include the leading-order eccentricity contribution to the tidal equations. The contribution of the stellar tide to the rates of change of orbital elements is:
\begin{eqnarray}
\dot a&=&-\frac{a}{9t_\mathrm{conv}}\frac{M_\mathrm{env}}{M_\star}\left(1+\frac{M_\mathrm{pl}}{M_\star}\right)\frac{M_\mathrm{pl}}{M_\star}\left(\frac{R_\star}{a}\right)^8 \label{eq:adotstar}\\
&&\times\left[2f_2+e^2\left(\frac{7}{8}f_1-10f_2+\frac{441}{8}f_3\right)\right]\nonumber\\
\dot e&=&-\frac{e}{36t_\mathrm{conv}}\frac{M_\mathrm{env}}{M_\star}\left(1+\frac{M_\mathrm{pl}}{M_\star}\right)\frac{M_\mathrm{pl}}{M_\star}\left(\frac{R_\star}{a}\right)^8 \label{eq:edotstar}\\
&&\times\left(\frac{5}{4}f_1-2f_2+\frac{147}{4}f_3\right).\nonumber\\
\end{eqnarray}
Here, $a$ and $e$ are the semi-major axis and eccentricity of the planet; $M_\star$, $M_\mathrm{env}$ and $M_\mathrm{pl}$ are the masses of the star, stellar envelope, and planet; $R_\star$ is the stellar radius; $t_\mathrm{conv}$ is the convective time-scale in the stellar envelope, given by
\begin{equation}\label{eq:tconv}
t_\mathrm{conv}=\left(\frac{M_\mathrm{env}R_\mathrm{env}^2}{\eta_\mathrm{F}L_\star}\right)^{1/3},
\end{equation}
$L_\star$ being the stellar luminosity, $R_\mathrm{env}$ the radius of the stellar envelope, and $\eta_\mathrm{F}$ a parameter of order unity. The coefficients $f_i$ in the tidal equations contain the frequency dependence of the components of the tidal forces, and are given by
\begin{equation}\label{eq:f}
f_i=f^\prime\mathrm{min}\left[1,\left(\frac{2\pi}{\sigma_i c_\mathrm{F} t_\mathrm{conv}}\right)^{\gamma_\mathrm{F}}\right].
\end{equation}
$\sigma_i=in$, $n$ being the mean motion, are the individual frequency components. $f^\prime$ and $c_\mathrm{F}$ are order-unity parameters, and $\gamma_\mathrm{F}$ an exponent determining the nature of the frequency dependence. Analytical and numerical work suggests that $\gamma_\mathrm{F}\approx 1-2$ \citep{Zahn77,GoldreichNicholson77,GoodmanOh97,Penev+07}. For compatibility with the results of \cite{VL09}, we adopt $\eta_\mathrm{F}=3$, $c_\mathrm{F}=1$, $f^\prime=9/2$ and $\gamma_\mathrm{F}=2$ as our fiducial parameters.

While we have chosen to study the effects of a particular tidal mechanism, we recognize that others may be at work. These could be dealt with by using the well-known $Q$ formalism, which has the advantage of putting the unknown strength of the tidal forces into the single parameter $Q^\prime_\star$. However, the value that this should take is not certain for giant stars. Studies of Main-Sequence planet hosts suggest $Q^\prime_\star$ in the range $10^5-10^{10}$ \citep[e.g.,][]{JacksonGreenbergBarnes08,PenevSasselov11}. However, even this poor calibration may not be applicable to giant stars with their very different structure: \cite{Nordhaus+10} find that their use of the Zahn formalism gives equivalent $Q^\prime_\star$ values of order $10^2-10^3$. Use of a $Q$ formalism with a Main-Sequence calibration will result in drastically weaker tides, while a calibration for giants must come from estimating the tidal strength per the assumed dissipation model, with little change if the turbulent dissipation model be adopted. Other models proposed for objects with convective envelopes, e.g., dissipation of inertial waves \citep[e.g.,][]{GoodmanLackner09}, will likely give much higher $Q$ values than the turbulent dissipation mechanism.

We also include, for the eccentric planets, a planetary tide. The dominant tidal mechanism in giant planet atmospheres is not known, with dynamical tide mechanisms such as gravity wave damping \citep*{Lubow+97} and inertial wave damping \citep{OgilvieLin04,Wu05a,Wu05b,GoodmanLackner09} having been proposed. In light of this uncertainty, we adopt the standard $Q$ formalism and a constant time lag of the tidal bulge \citep{MatsumuraPealeRasio10}. This affects the evolution of the orbital elements as
\begin{eqnarray}
\dot a&=&6\frac{M_\star}{M_\mathrm{pl}}\Delta_\mathrm{pl}\left(\frac{R_\mathrm{pl}}{a}\right)^4R_\mathrm{pl}\\
&&\times\left[\left(1+\frac{27}{2}e^2\right)\Omega_\mathrm{pl}-\left(1-23e^2\right)n\right]\nonumber\\
\dot e&=&27\frac{M_\star}{M_\mathrm{pl}}\Delta_\mathrm{pl}\left(\frac{R_\mathrm{pl}}{a}\right)^5e\\
&&\times\left(\frac{11}{18}\Omega_\mathrm{pl}-n\right),\nonumber
\end{eqnarray}
with $\Delta_\mathrm{pl}=3n/2Q^\prime_\mathrm{pl}|2\Omega_\mathrm{pl}-n|$ being the product of the mean motion, the lag time, and the planetary Love number. A constant time lag entails varying $Q^\prime_\mathrm{pl}$; our $Q^\prime_\mathrm{pl}$ values quoted below are normalized to $0.05$\,AU for comparison with planets around more commonly studied Main Sequence stars. We test a range of $Q^\prime_\mathrm{pl}$ values, from $10^2$ to $10^{10}$ for giant planets and $10^0$ to $10^3$ for terrestrials; however, we find that in all cases the planetary tide is so weak as to have no discernible effect on the orbital evolution. The planetary angular momentum $\Omega_\mathrm{pl}$ also evolves:
\begin{eqnarray}
\dot\Omega_\mathrm{pl}&=&3\Delta_\mathrm{pl}\frac{M_\star^2}{M_\mathrm{pl}\left(M_\star+M_\mathrm{pl}\right)\alpha_\mathrm{pl}^2} R_\mathrm{pl}^3 n\\
&&\times\left[\left(1+\frac{27}{2}e^2\right)n-\left(1+\frac{15}{2}e^2\right)\Omega_\mathrm{pl}\right].\nonumber
\end{eqnarray}
 The star is assumed to be non-rotating, which is likely a reasonable approximation as the spin rates of giants are very low: rotational velocities of RGB stars are on average $ v\sin i \le 2 \mathrm{km\,s}^{-1}$ \citep[see e.g.][]{deMedeiros+99}, with rapid rotators representing less than 2\% of a sample of 1300 K~giants \citep{Carlberg+11}. Conservation of angular momentum should ensure that stars during the AGB phase are slow rotators as well; for they will spin down after the horizontal branch as their radii expand.

Finally, the planet's orbit can expand as the star loses mass. All expansion is assumed to be adiabatic, with the eccentricity remaining constant and the semi-major axis varying by
\begin{equation}
\dot a=-a\frac{\dot M_\star+\dot M_\mathrm{pl}}{M_\star+M_\mathrm{pl}},
\end{equation}
although we do not include the rate of change of the planet's mass due to either mass accretion or evaporation \citep[see][]{VL09} because it is negligible compared to the stellar mass loss term during the AGB phase. The assumption of adiabaticity is justified so long as the time-scale for mass loss is much longer than the Keplerian time-scale. This criterion was parameterized by \cite{Veras+11} as
\begin{equation}
\Psi=\frac{1}{2\pi}
\frac{\dot M}{\mathrm{M}_\odot\mathrm{\,yr}^{-1}}
\left(\frac{a}{\mathrm{AU}}\right)^{3/2}
\left(\frac{M_\star}{\mathrm{M}_\odot}\right)^{-3/2},
\end{equation}
finding that substantial deviations from adiabaticity could arise for $\Psi$ as small as $0.02$, this causing eccentricity variations of order 0.1 if maintained for several thousands of years. For our stellar models, taking a planet at 5\,AU, the maximum $\Psi$ is $3\times 10^{-3}$, attained only briefly, so we expect that the adiabatic approximation will hold.

We neglect any drag forces from the strong stellar wind, shown by \cite{DL98} to reduce a planet's semi-major axis by a mere 1 per cent over the AGB lifetime.

\section{Stellar models}

\label{sec:models}

As described previously, the chosen parameterization to calculate the tidal
forces involves the dissipation of energy in the convective
envelope. In order to calculate its effect on the orbital evolution of the
planet a number of parameters related
to the stellar interior such as the mass and radius of the convective
envelope are needed. A good parameterization of the mass loss during this phase is especially critical as well, because its time-scale is linked to the
stellar structure and because it is a determining factor in the planet's
orbital evolution.

The internal structure of an AGB star is complex, especially during
the thermally-pulsing stage when strong time-dependency occurs and when
most of the mass loss takes place. During this phase two
nuclear burning shells, the inner thermally
unstable and burning helium, the outer composed of hydrogen, surround a degenerate carbon--oxygen
core \citep{Schwarzschild65}. During each thermal pulse the inner
edge of the convective envelope moves inwards, close to the H-burning shell, dredging up fresh material to
the surface. Variations in the surface luminosity and radius are
produced as a consequence of the thermal readjustments taking place
during the thermal pulses \citep[see, e.g.,][]{KarakasLattanzio07}.

These thermal pulses responsible for the modulations in the stellar
radius are also the drivers of the mass loss through the generation
of shocks in the stellar envelope that allow the nucleation of dust
grains \citep[see, e.g.,][and references therein]{Bowen88}. As grains
nucleate and grow they experience
the force exerted by the stellar radiation pressure and thus are
accelerated away from the star. The momentum coupling between gas and dust then drives the mass outflow.

We have used the temporal evolution of mass loss provided by the stellar
evolutionary models of \cite{VassiliadisWood93}; see also \cite{Villaver+02,Villaver+12} for a discussion on the
mass-loss in these models. For consistency, the stellar structure is
derived from the same models using the following approximations.
First, the envelope mass $M_\mathrm{env}$ is approximated by $M_{\star}-M_\mathrm{c}$,
where $M_{\star}$ is the star's total mass, and $M_\mathrm{c}$, the core mass, has been taken as its value at the first
thermal pulse from the parameterization, as a function of the initial mass and metallicity, given by \cite{Karakas+02}. The core mass is expected to
increase slightly due to H--burning during the inter-pulse periods. This increase
in mass, although fundamental in problems devoted to stellar evolution,
is negligible for our purposes, being a few hundredths of a Solar mass. The use of the initial core mass, so that the envelope mass never goes to zero, also avoids numerical problems when integrating the equations of motion (Equations \ref{eq:adotstar}, \ref{eq:edotstar} and \ref{eq:f}). The envelope mass is then made to decrease
with the stellar mass.
Second, we have taken the radius of the convective envelope to be the
radius of the star. The region interior to the H-burning shell is
similar in size to a white dwarf, with almost the entire stellar volume taken up by the
convective envelope, and the radius at the base
of the convective envelope being just $\approx 10^{-4} R_\star$ \citep[see, e.g.,][]{Marigo+98}. This approximation thus has a negligible effect when it enters in the estimate of the
convective time-scale in Equation\,(\ref{eq:tconv}).

\section{Numerical studies}

\label{sec:numerics}

\subsection{Planets on circular orbits}

\begin{figure}
  \includegraphics[width=.5\textwidth]{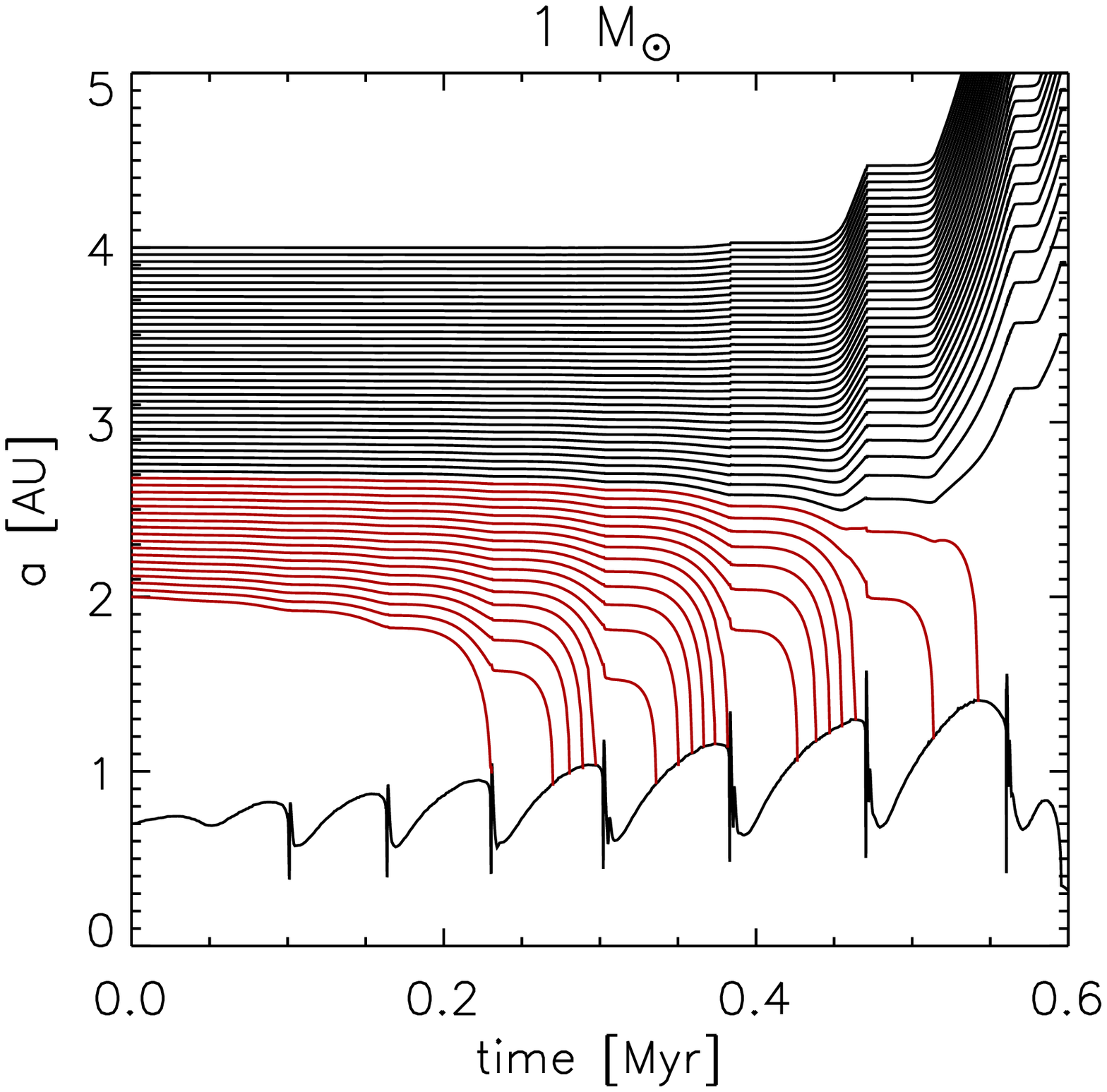}
  \includegraphics[width=.5\textwidth]{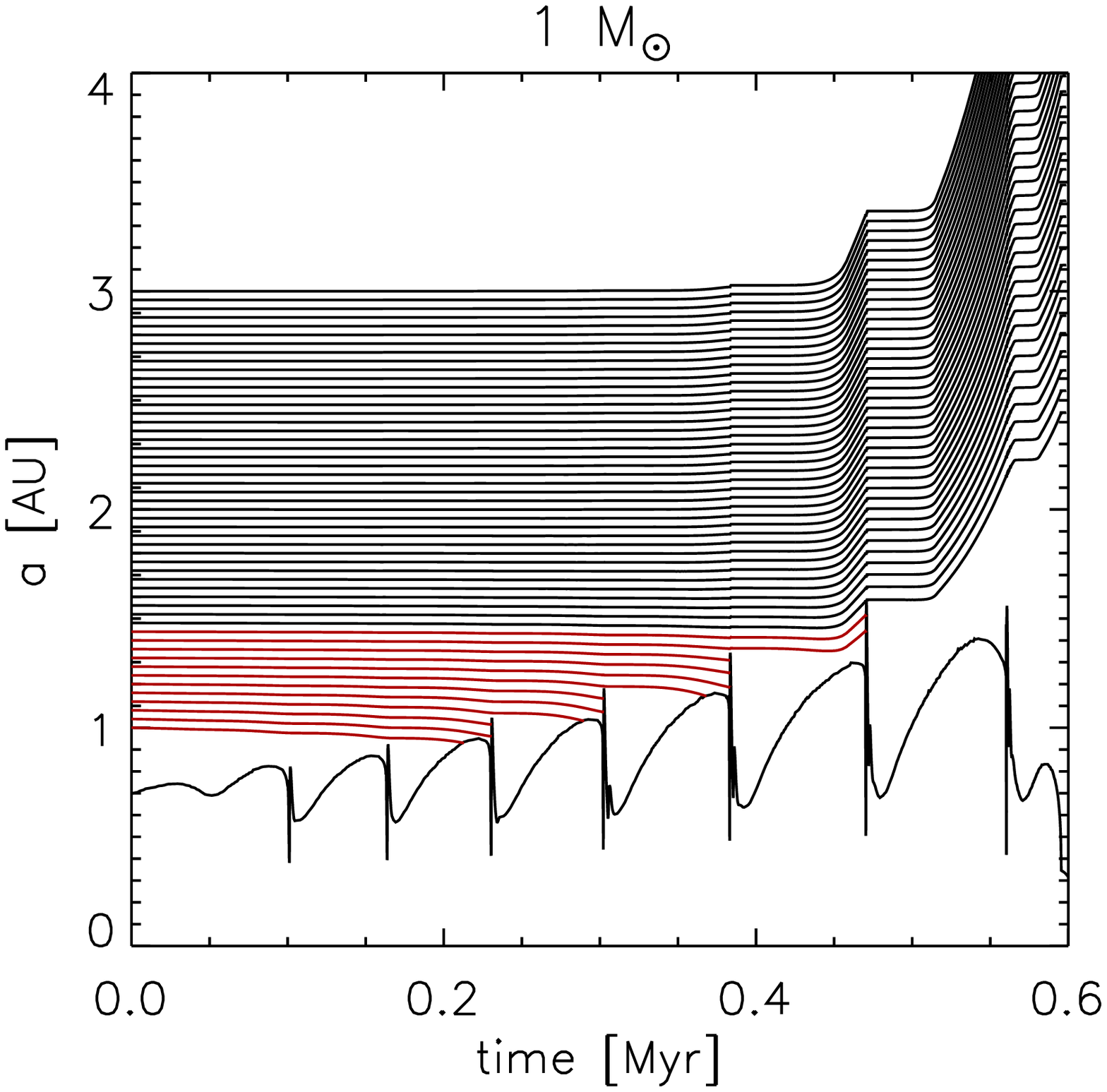}
  \caption{Orbital evolution of planets initially at different semi-major axes around a $1\mathrm{\,M}_\odot$ AGB star, under the effects of tidal forces and stellar mass-loss. Planets in the top panel are of Jupiter's mass; those in the bottom are of Earth's mass. Black lines show planets that survive; grey lines (red on-line) show planets that strike the stellar envelope.}
  \label{fig:1msol}
\end{figure}

\begin{figure}
  \includegraphics[width=.5\textwidth]{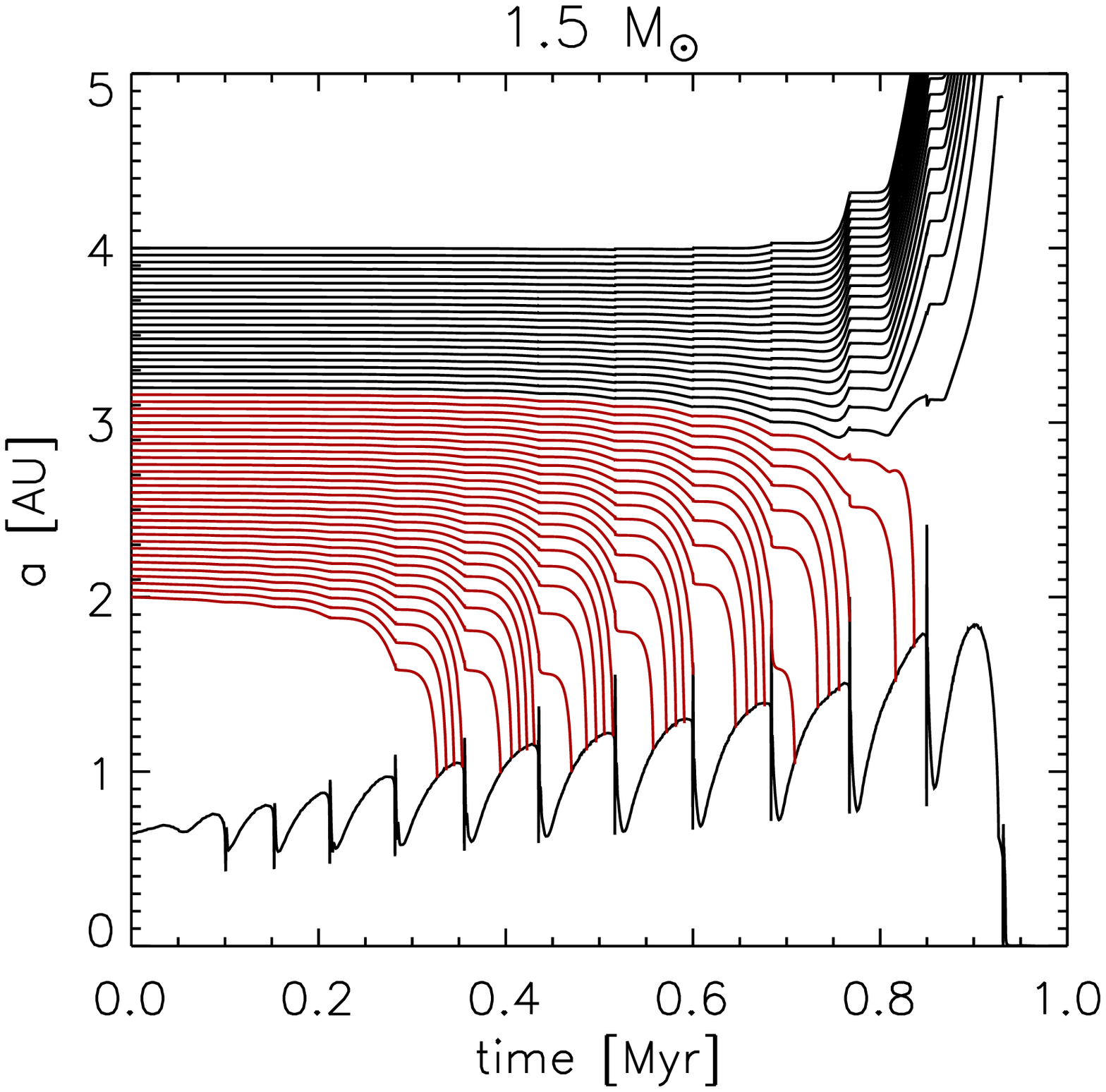}
  \includegraphics[width=.5\textwidth]{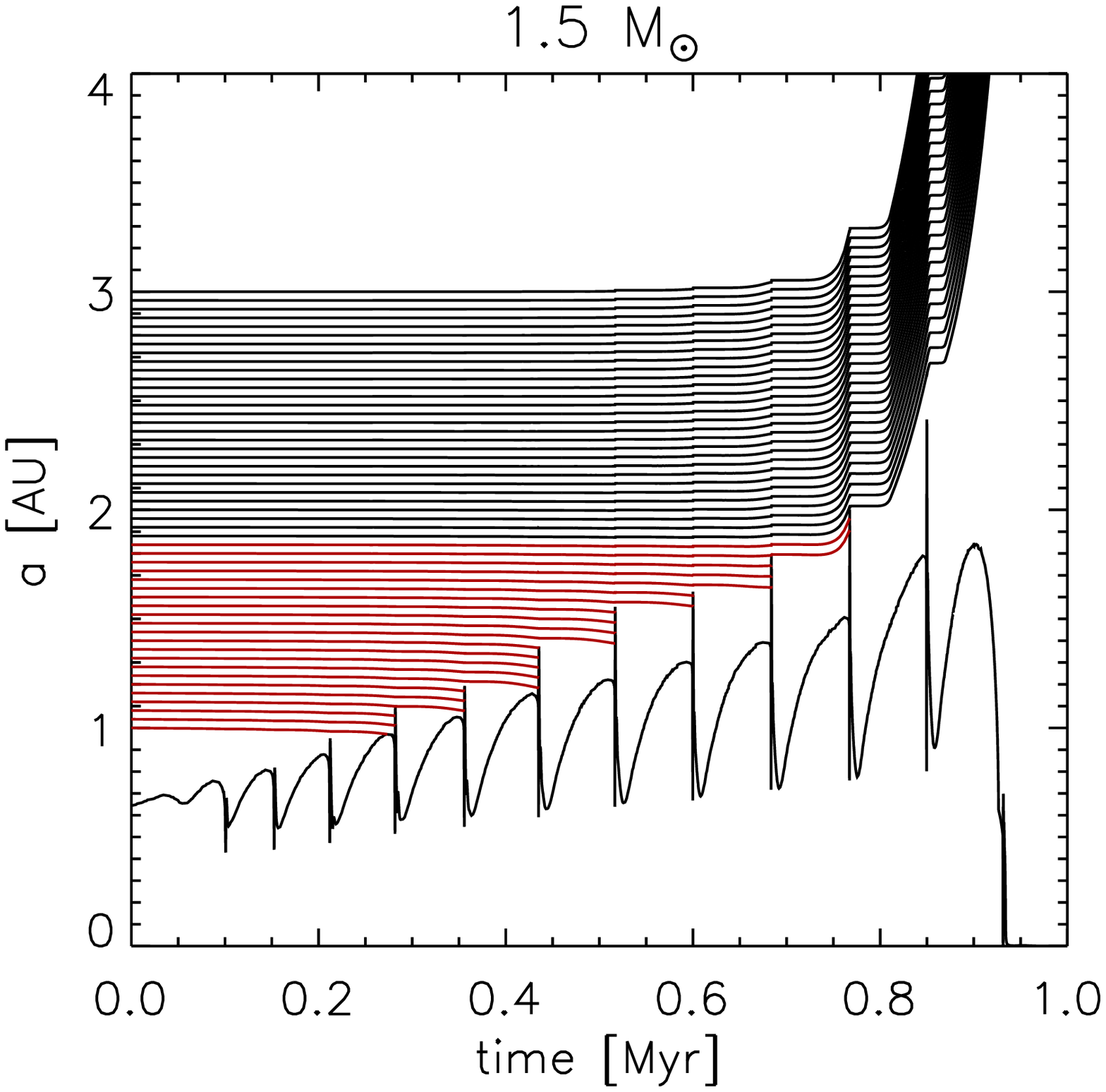}
  \caption{As Figure~\ref{fig:1msol} but for a $1.5\mathrm{\,M}_\odot$ star. Note here, and also in subsequent plots, the different axis scales.}
  \label{fig:1.5msol}
\end{figure}

\begin{figure}
  \includegraphics[width=.5\textwidth]{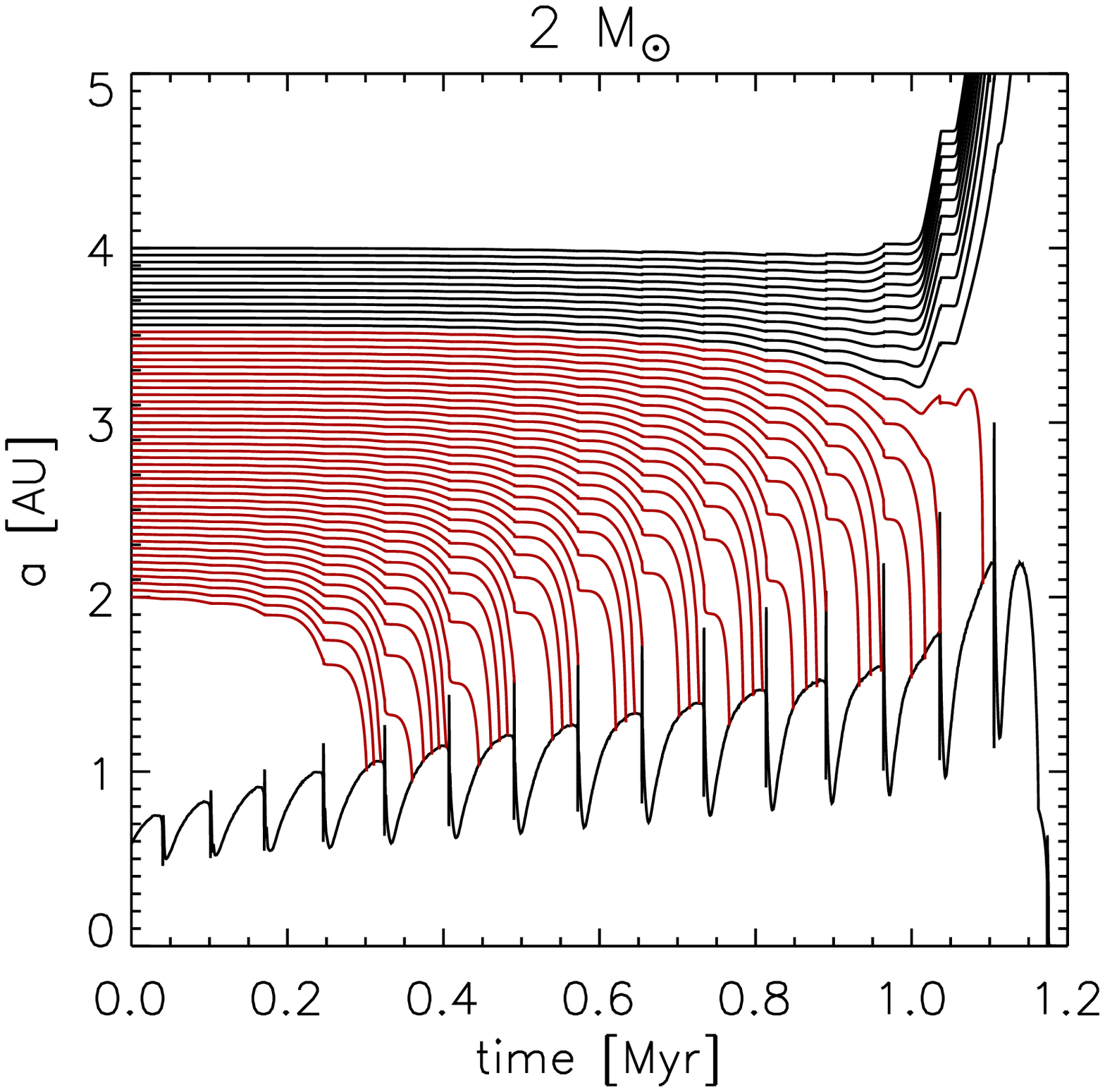}
  \includegraphics[width=.5\textwidth]{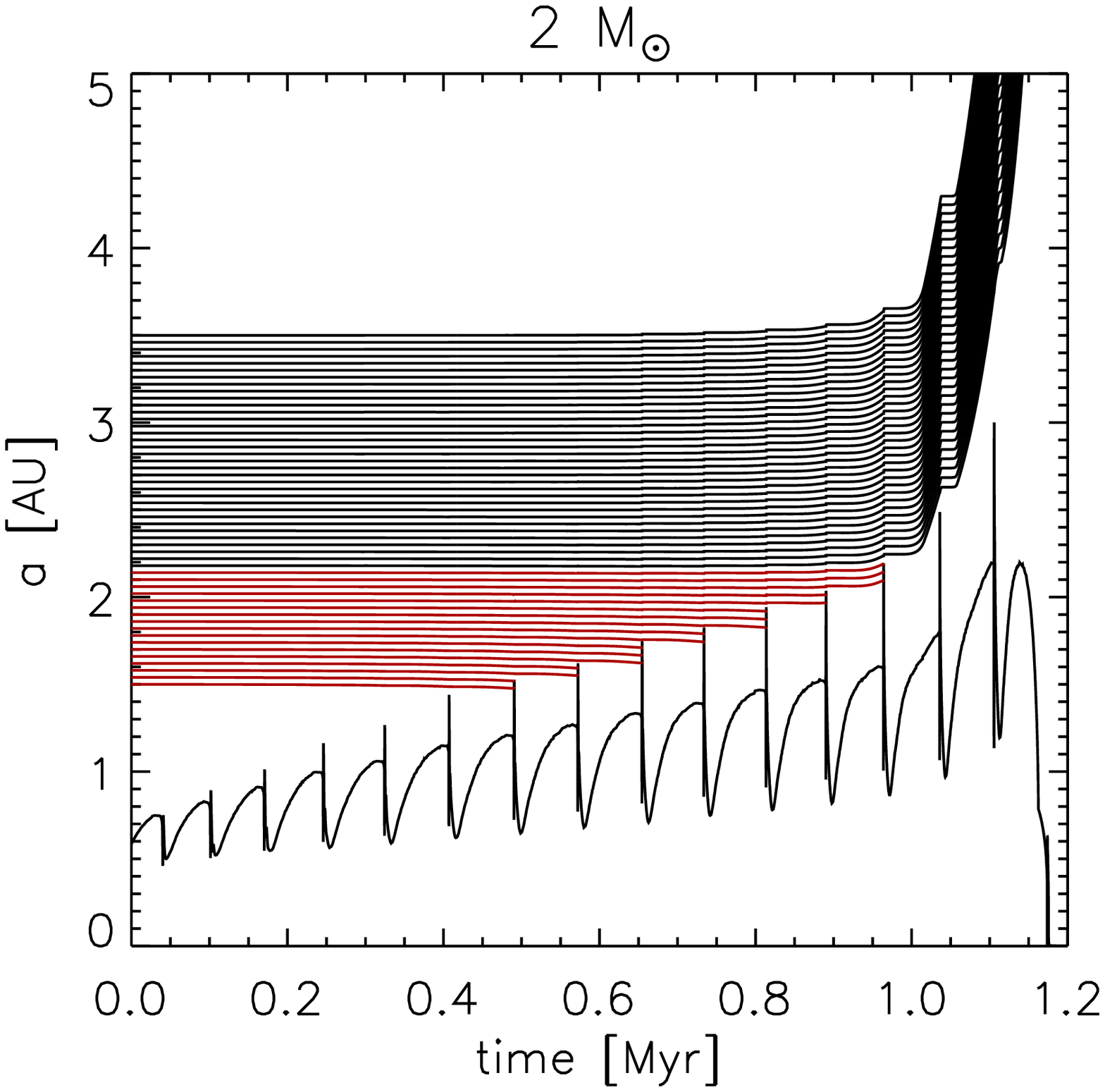}
  \caption{As Figure~\ref{fig:1msol} but for a $2\mathrm{\,M}_\odot$ star.}
  \label{fig:2msol}
\end{figure}

\begin{figure}
  \includegraphics[width=.5\textwidth]{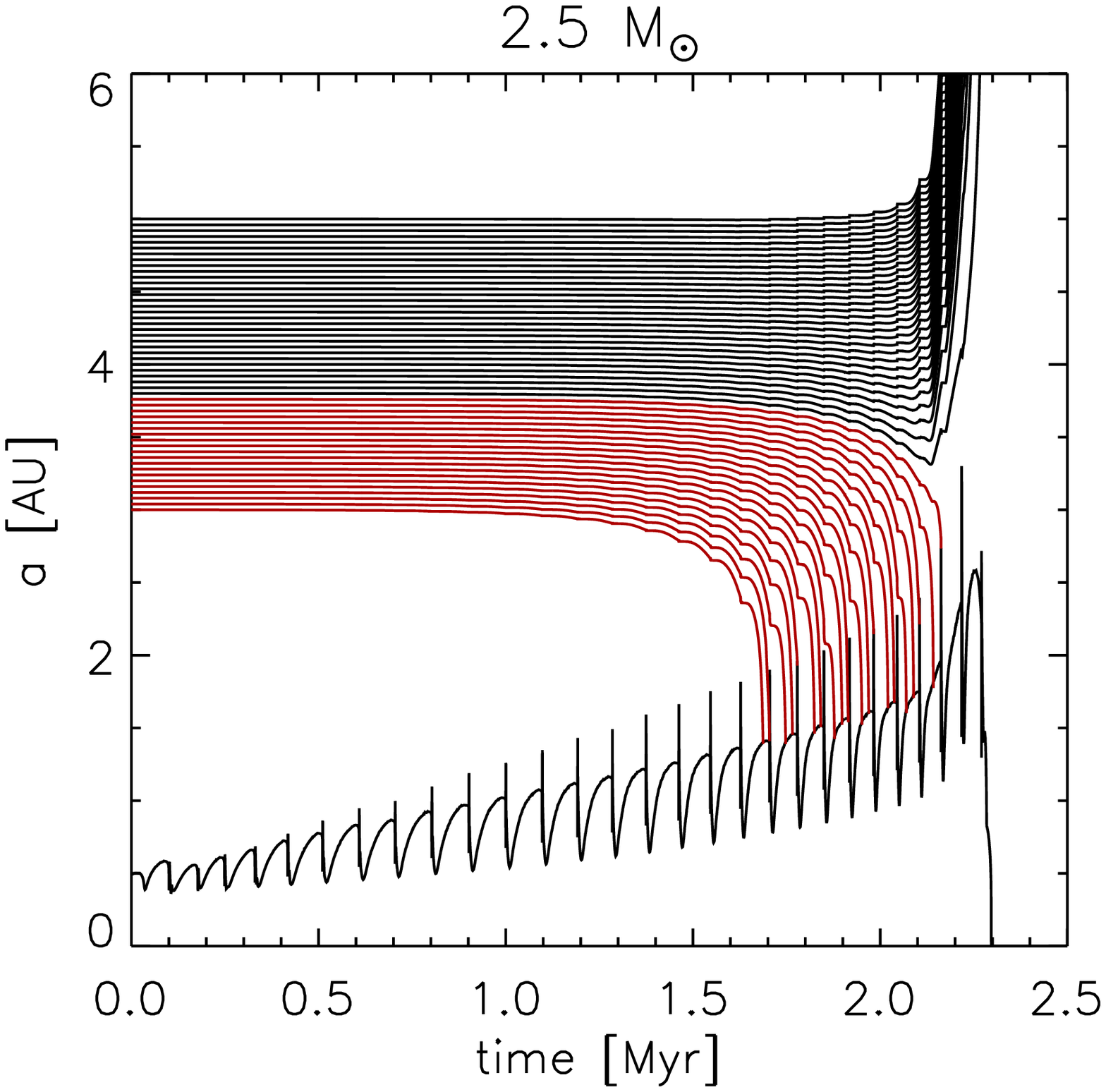}
  \includegraphics[width=.5\textwidth]{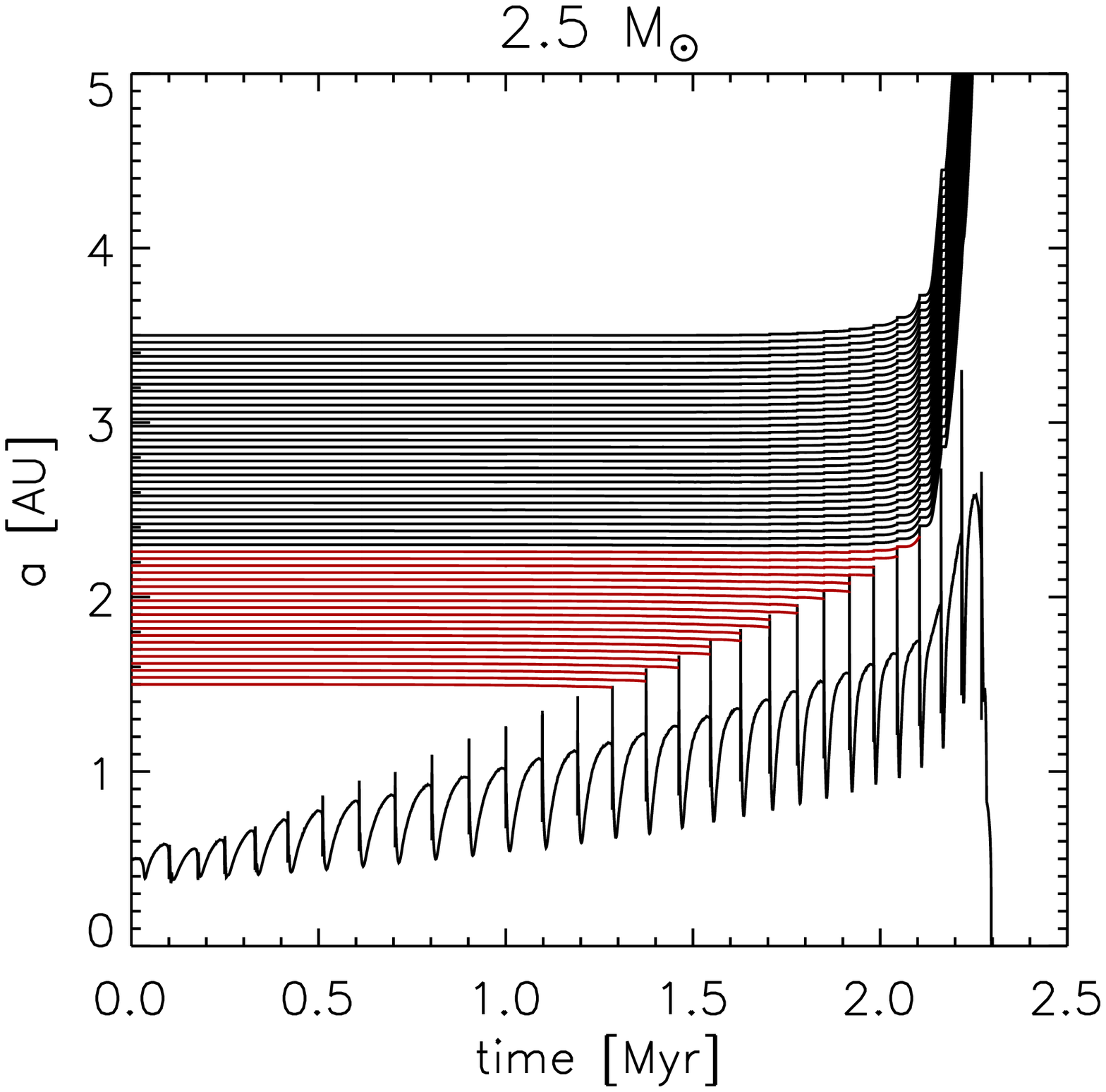}
  \caption{As Figure~\ref{fig:1msol} but for a $2.5\mathrm{\,M}_\odot$ star.}
  \label{fig:2.5msol}
\end{figure}

\begin{figure}
  \includegraphics[width=.5\textwidth]{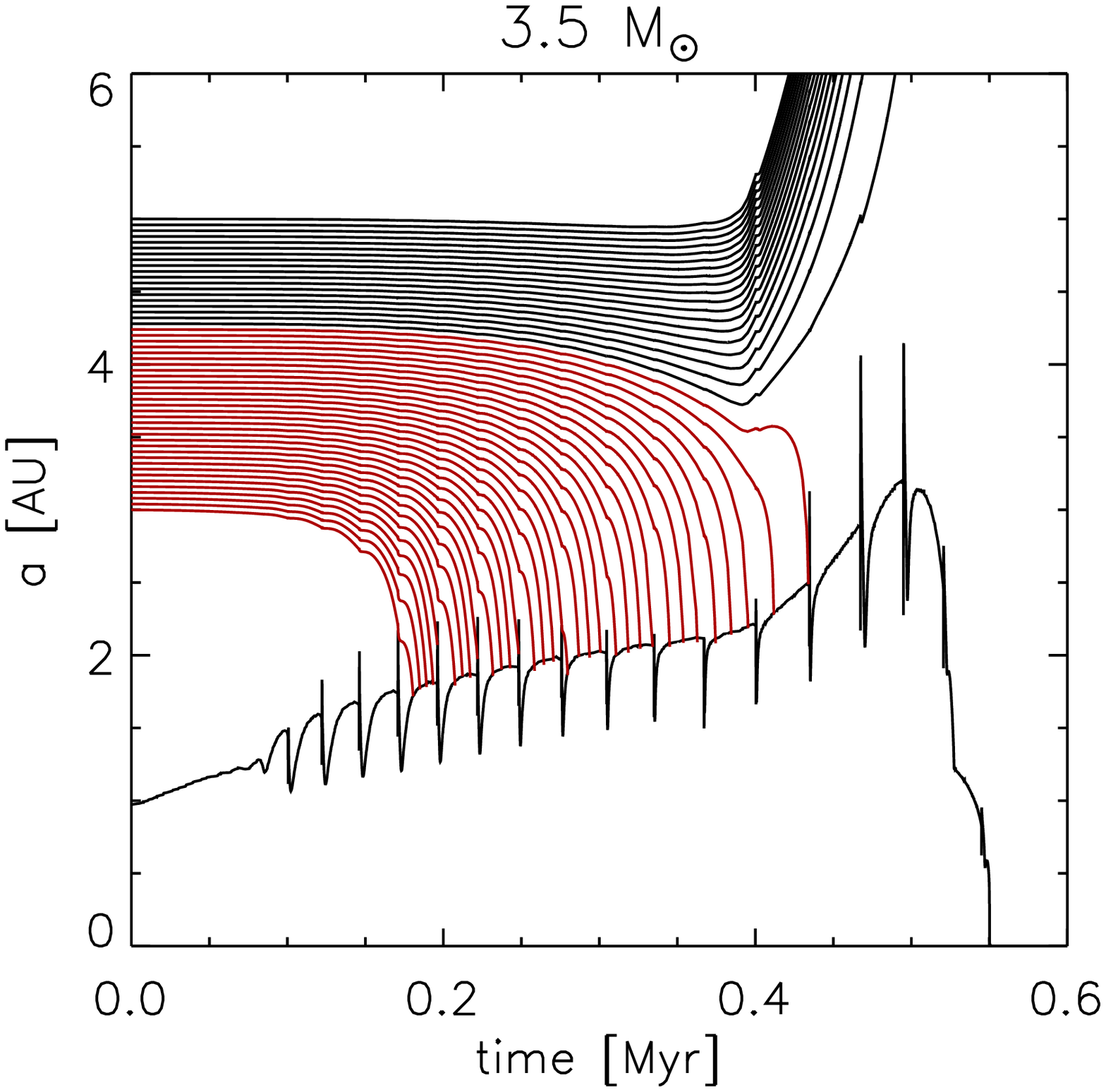}
  \includegraphics[width=.5\textwidth]{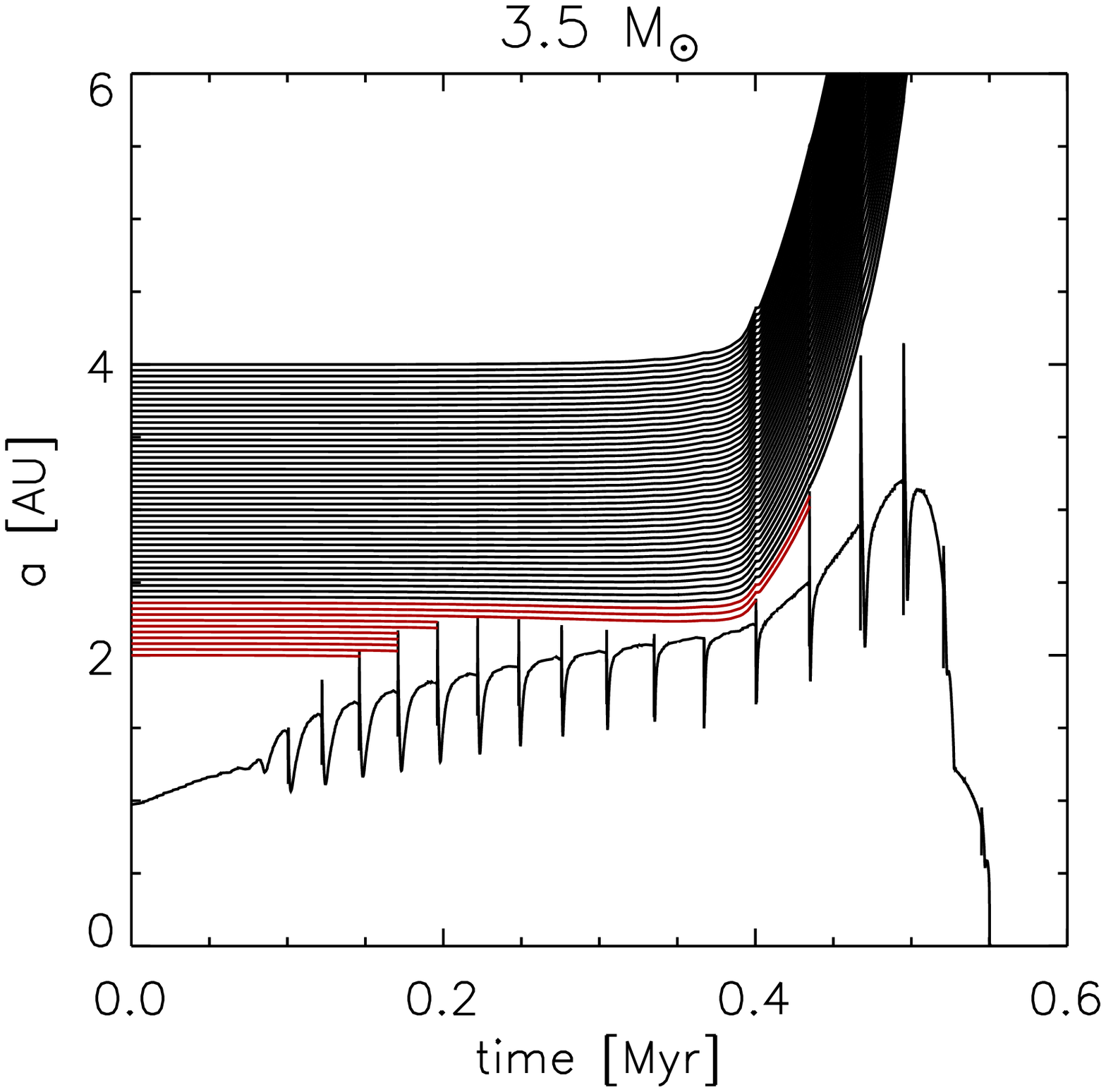}
  \caption{As Figure~\ref{fig:1msol} but for a $3.5\mathrm{\,M}_\odot$ star.}
  \label{fig:3.5msol}
\end{figure}

\begin{figure}
  \includegraphics[width=.5\textwidth]{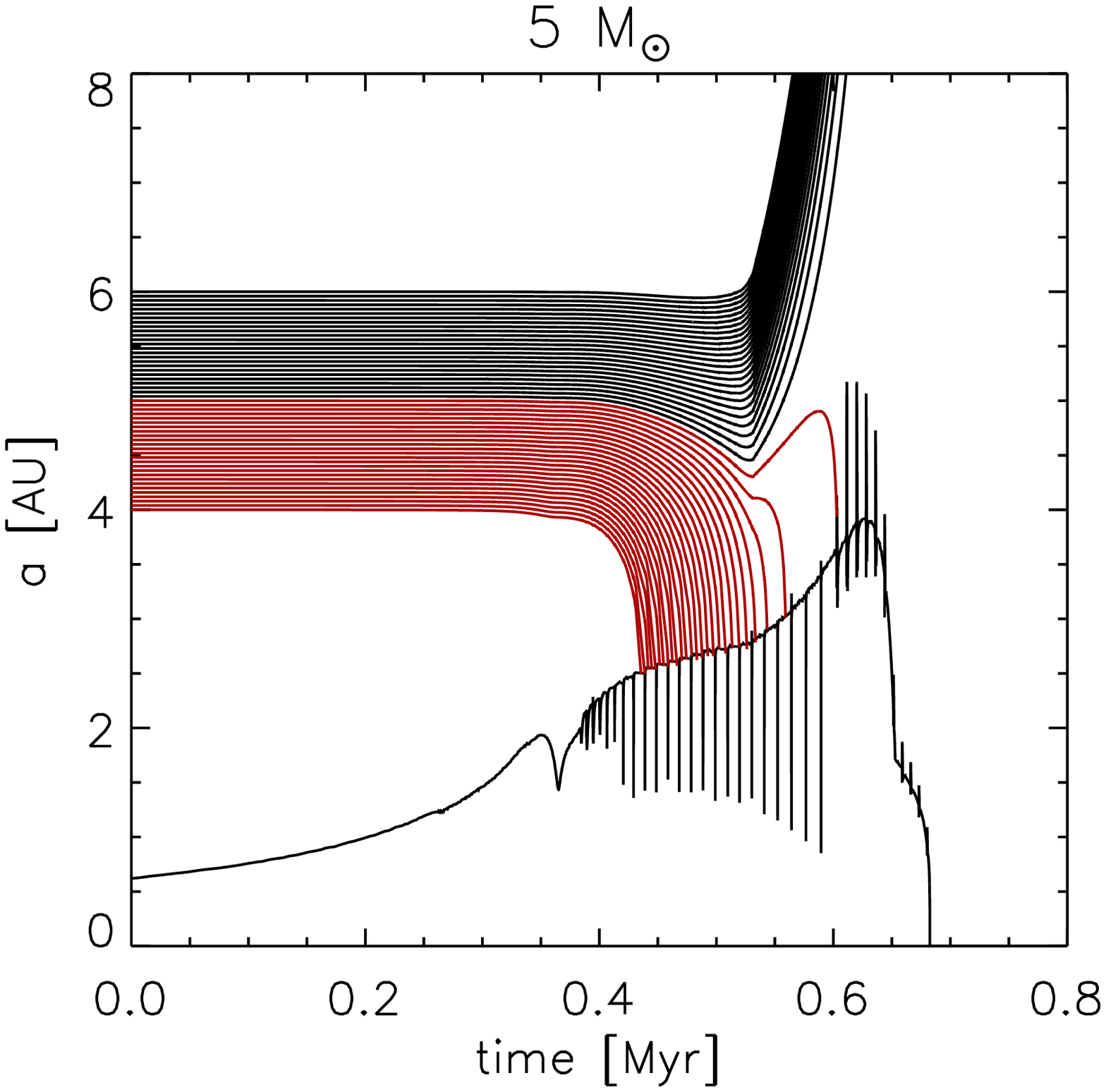}
  \includegraphics[width=.5\textwidth]{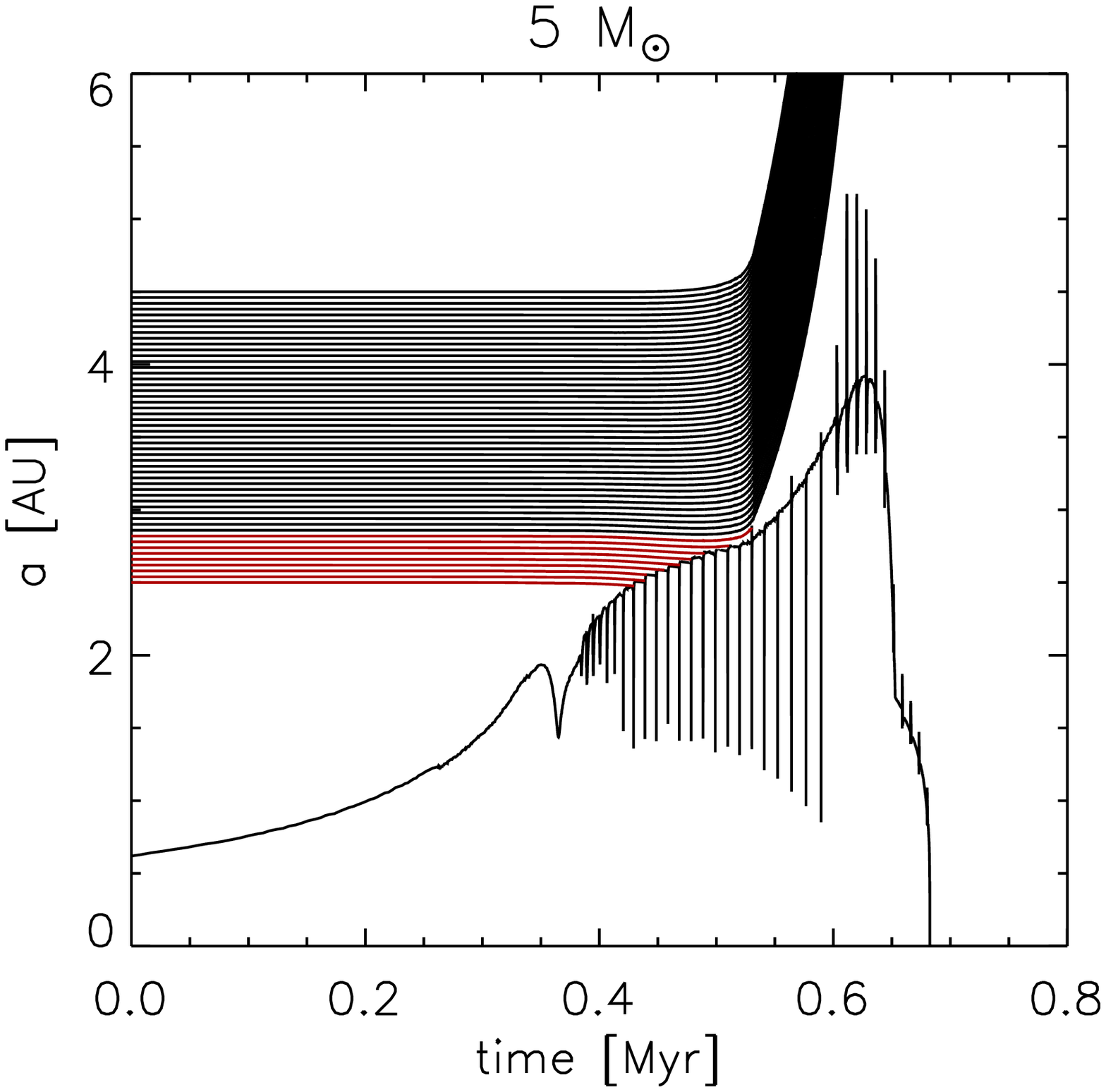}
  \caption{As Figure~\ref{fig:1msol} but for a $5\mathrm{\,M}_\odot$ star.}
  \label{fig:5msol}
\end{figure}

We first considered the evolution of planets initially on circular orbits about the stars of initial mass $1$, $1.5$, $2$, $2.5$, $3.5$ and $5\mathrm{\,M}_\odot$, for the duration of the stars' thermally pulsing AGB lifetime. Note that the ``initial'' mass is that at the beginning of the AGB; this may have been reduced below the MS mass if mass loss has already occurred on the RGB. We considered planets of Terrestrial ($1\mathrm{\,M}_\oplus$), Neptunian ($17.1\mathrm{\,M}_\oplus$) and Jovian ($318\mathrm{\,M}_\oplus$) mass. The tidal equations in Section~\ref{sec:eqns}, including orbital expansion from mass loss, were integrated with a Runge--Kutta integrator \citep{NumRec}. The integration was stopped if the planet at any time entered the stellar envelope. The lifetime of planets having entered the envelope is very short, and despite the brevity of some of the radial pulses we expect that the orbits of such planets will decay rapidly. Loss of orbital energy is very fast and the time-scale can be as short as weeks \citep{NordhausBlackman06}. The question of the potential final fate of planets engulfed in the stellar envelope is not straightforward. \cite{VL07} estimated that a planet up to $15\mathrm{\,M_J}$ can be evaporated inside the envelope of an AGB star with a main-sequence mass of $1\mathrm{\,M}_\odot$, with this mass limit increasing and reaching well into the stellar regime ($\sim120\mathrm{\,M_J}$, $0.11\mathrm{\,M}_\odot$) if the planet (or brown dwarf) is engulfed in the envelope of a $5\mathrm{\,M}_\odot$ AGB star. However, as has
been argued several times \citep{Villaver11,Villaver12} the value of this maximum mass is very uncertain because it depends on several factors, such as the efficiency of stellar envelope ejection, which are largely unknown. Furthermore, to survive the Common Envelope phase, a planet must be able to supply enough of its orbital energy to the stellar envelope to unbind the envelope before it spirals down to the Roche radius, at which the planet is tidally disrupted. \cite{Nordhaus+10} estimated the minimum mass for this to happen at just under 10 Jovian masses for a $1\mathrm{\,M}_\odot$ star, much larger than the planets we are considering. Note that the Roche radius is very small compared to the radius of the star, so planetary disruption following passage of the Roche radius will only occur if the planet has already entered the envelope.

\begin{figure}
  \includegraphics[width=.5\textwidth]{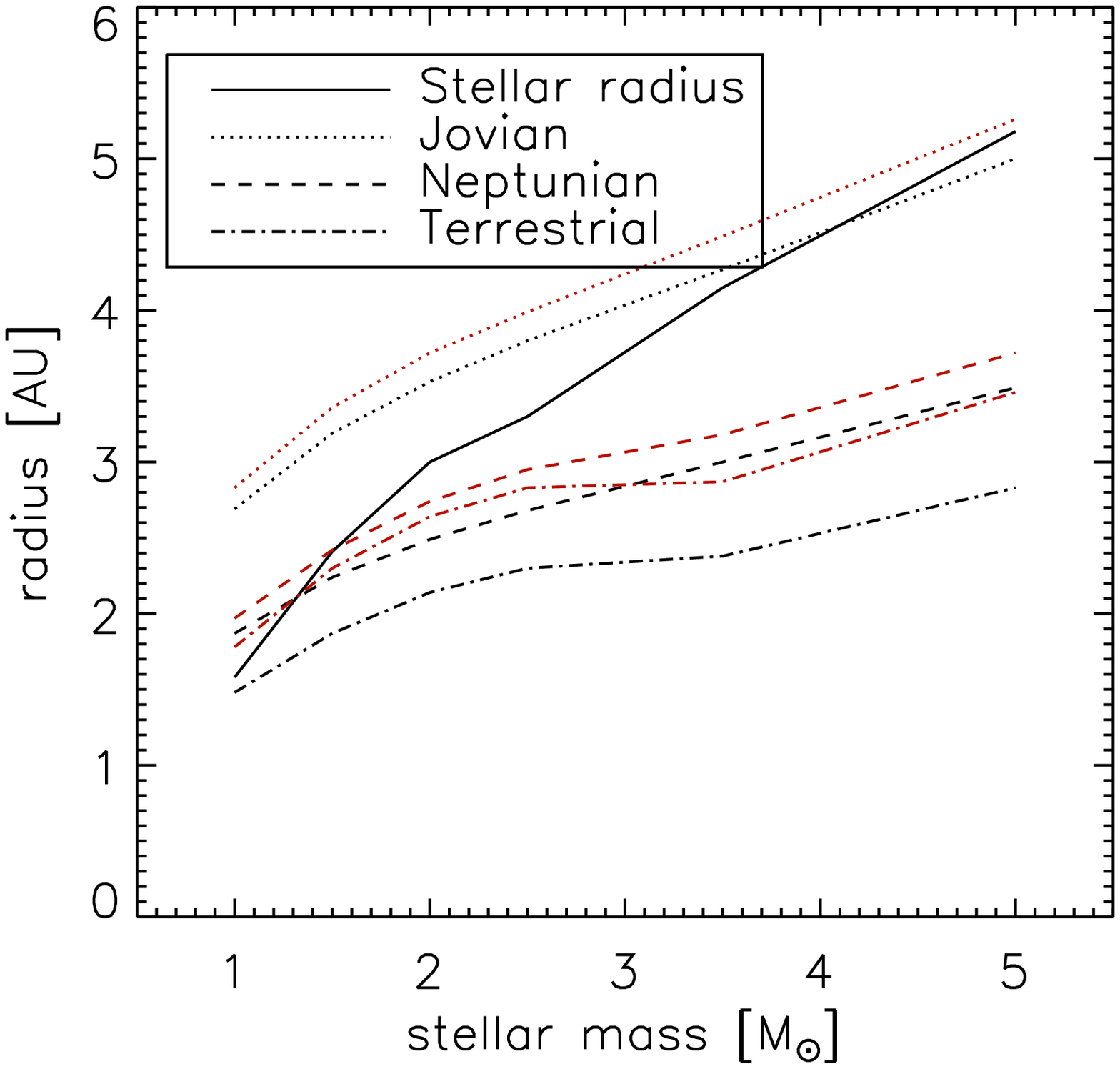}
  \caption{Maximum radius of the star on the AGB (solid line) and the orbital radii of the initially most distant planet engulfed by the star, as a function of stellar mass. Black lines show planets on circular orbits; grey lines (red in the on-line version), planets initially on eccentric orbits.}
  \label{fig:morituri}
\end{figure}

In Figures~\ref{fig:1msol} to~\ref{fig:5msol} we show the semi-major axis evolution of planets of Jovian and Terrestrial mass around the different stars; the stellar radius is also shown. The evolution of the planets orbiting each star displays the expected conflict between orbital expansion due to stellar mass loss and orbital contraction due to tidal effects. Towards the end of each thermal pulse, tidal decay accelerates as the stellar radius expands, but may then be reversed as the star loses mass at the end of the pulse. Some planets undergo several cycles of motion inwards and outwards before finally colliding with the star or escaping to safe radii. The evolution of the planet's orbit depends only on its mass and initial semi-major axis, with all planets interior to a critical radius being engulfed and all exterior avoiding engulfment. For brevity, we refer to the engulfed planet with the largest initial semi-major axis as the \emph{moriturus ultimus.} The orbit of the moriturus ultimus is at several AU and is larger for the more massive stars and more massive planets. These radii are plotted in Figure~\ref{fig:morituri} as black lines.

For the Jovian planets, tidal decay can be significant, since tidal forces are strong for massive planets. Around lower mass stars, tidal forces can pull in planets that have initial orbits far larger than the maximum stellar radius. This is not the case, however, for highest mass star ($5\mathrm{\,M}_\odot$), where the different pattern of stellar radius evolution and mass loss, and smaller planet:star mass ratio, means that some planets which begin interior to the maximum stellar radius will be moved beyond it by mass loss before this radius is attained, and be thus saved from entering the stellar envelope.

For the Terrestrial planets, tidal forces are very weak, and the evolution is dominated by the expansion due to mass-loss. This is in all cases sufficient to push some planets that begin interior to the maximum stellar radius out to safety. This is particularly noticeable in for the $5\mathrm{\,M}_\odot$  progenitor, which expands to over 5 AU, planets beyond 2.8 AU however being safe. The entry of these planets into the stellar envelope typically occurs when the stellar radius suddenly swells to engulf them.

Neptune-mass planets, as shown in Figure~\ref{fig:morituri}, display behaviour intermediate between Jovian and Terrestrial planets, as is to be expected. The semi-major axis of the moriturus ultimus for each star is slightly larger than for the Terrestrial planet, and in all cases save the $1\mathrm{\,M}_\odot$ star it lies within the maximum stellar radius.

\subsection{Inclusion of orbital eccentricity}

We now turn to consider the effects of moderate planetary eccentricities on the evolution of the planets' orbits around AGB stars. Because the ratio of the planetary radius to the orbital radius is so small, the eccentricity evolution is governed entirely by the stellar tide, in contrast to most cases of planets orbiting Main Sequence stars where both planetary and stellar tides are effective \citep[e.g.,][]{JacksonGreenbergBarnes08,Hansen10}, a distinction pointed out by \cite{Nordhaus+10}. Here, we include the lowest-order terms in eccentricity in the equations for the stellar tides as described in Section~\ref{sec:eqns}, and also include for completeness the lowest-order contribution to the planetary tide as described in Section\,\ref{sec:eqns}. Because of the dependence of the tidal response on forcing frequency, an expansion to arbitrarily high eccentricity using the Eggleton formalism \citep*{EggletonKiselevaHut98} is inconsistent with our assumed tidal model. We therefore restrict attention to planets with an initial eccentricity of 0.2, for which the lowest order equations were found to be accurate when compared with the complete \cite{EggletonKiselevaHut98} equations for the constant time-lag model.

\begin{figure}
  \hspace{-1cm}\includegraphics[width=.5\textwidth]{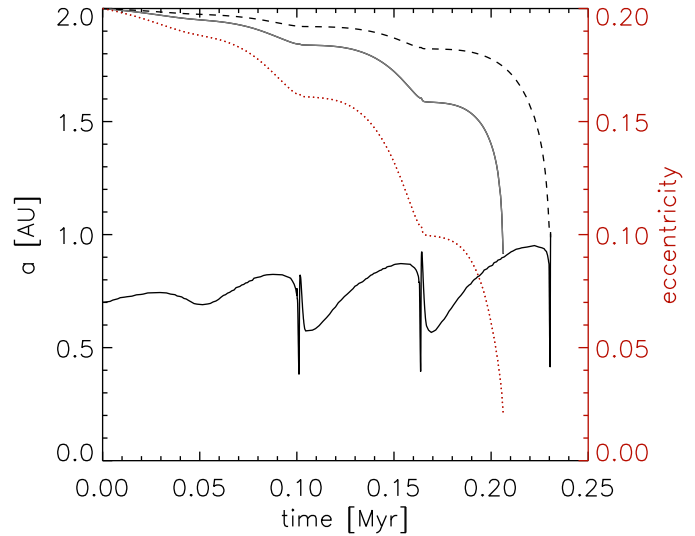}\\
  \hspace*{-1cm}\includegraphics[width=.5\textwidth]{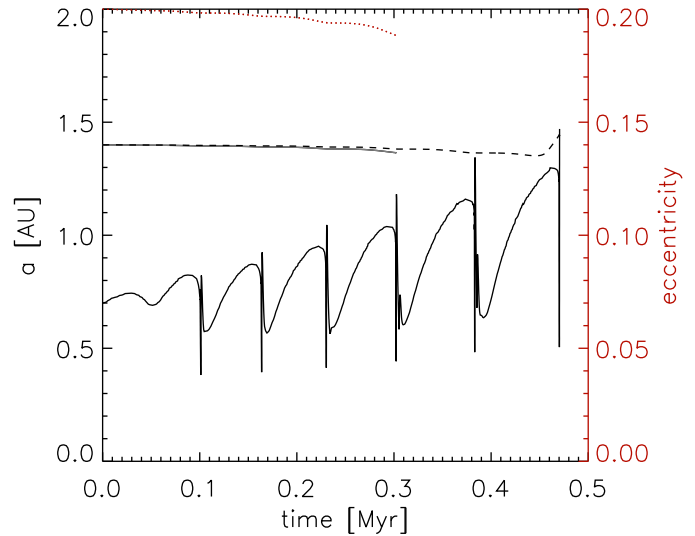}
  \caption{Top: Tidal evolution of Jovian planets around a $1\mathrm{\,M}_\odot$ AGB star. The solid black line shows the stellar radius; the dashed black line the semi-major axis of a planet on a circular orbit. The solid grey line shows the semi-major axis of planets with initial eccentricity of $0.2$. There are two lines here, with $Q_\mathrm{pl}^\prime=10^2$ and $10^{10}$, but they are indistinguishable. The dotted grey line (red on-line) and right-hand axis show the evolution of the eccentricity. Bottom: the same, for Terrestrial planets. Here the two eccentric planets have quality factors 1 and 1000, but again their evolution is indistinguishable.}
  \label{fig:ecc}
\end{figure}

The tidal evolution of example planets orbiting a star of $1\mathrm{\,M}_\odot$ is shown in Figure~\ref{fig:ecc}. That of Jovian planets is shown in the top panel. The eccentricity and semi-major axis damp on similar time-scales, and the inclusion of eccentricity shortens the lifetime of the Jovian planet by about 10 per cent compared to the lifetime of a non-eccentric planet. For other initial conditions, reductions of over 20 per cent were seen. The eccentricity has decayed by a factor of 10 by the time the planet hits the stellar envelope. The evolution of planets with quality factors $Q_\mathrm{pl}^\prime=10^2$ and $10^{10}$ was considered, comfortably bracketing estimated $Q$ values for gas giant planets, but planetary tides are so weak that they are indistinguishable. The bottom panel of Figure~\ref{fig:ecc} shows the evolution of Terrestrial planets, with the two $Q_\mathrm{pl}^\prime$ values considered, $10^0$ and $10^3$, again being indistinguishable. Here, stellar tides too are weak: the eccentricity barely decays, the inclusion of the eccentricity only slightly enhances the semi-major axis decay, and yet the lifetime of the planet is reduced by a third. This is because the planet's pericentre takes it inside the envelope during an earlier radial pulse than the one which swallows the planet on a circular orbit. For other initial conditions, the planet's lifetime could be halved by giving it an eccentricity of $0.2$.

\begin{figure}
  \includegraphics[width=.5\textwidth]{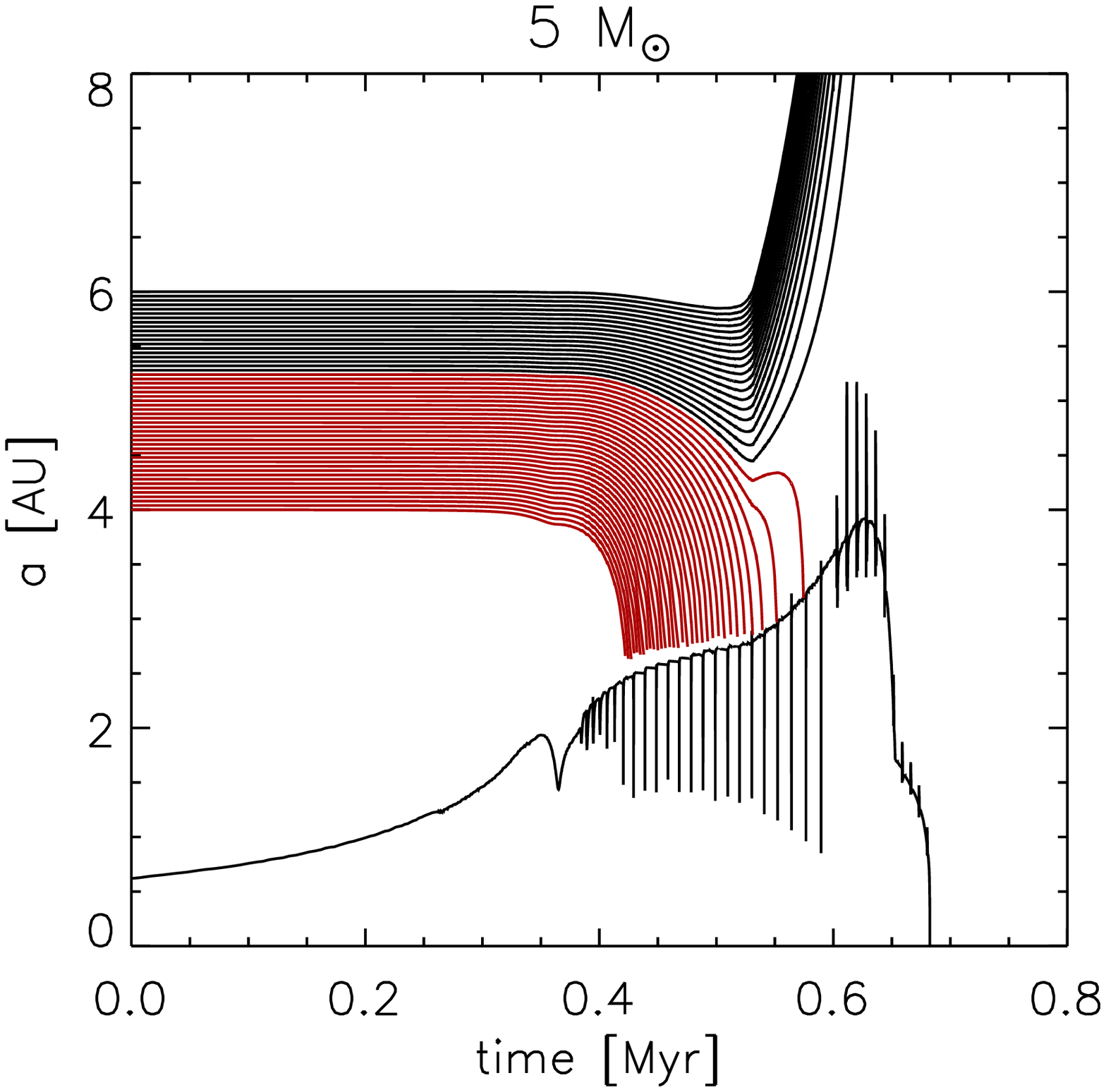}
  \includegraphics[width=.5\textwidth]{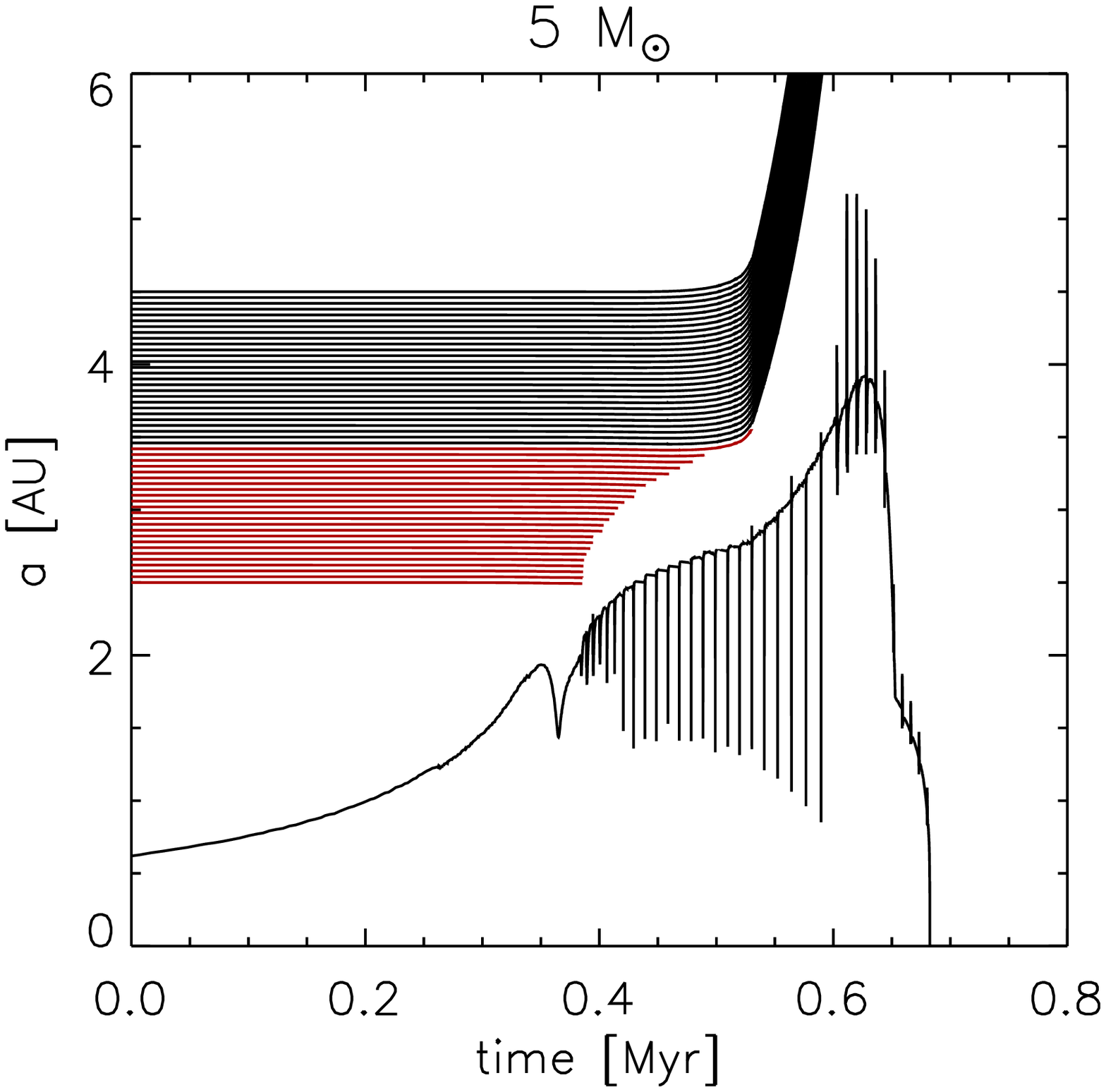}
  \caption{Semi-major axis evolution of eccentric planets around a $5\mathrm{\,M}_\odot$ star. Initial planetary eccentricity is $0.2$. Top: Jovian planets. Here the eccentricity decays to a small value prior to engulfment. Bottom: Terrestrial planets. Here there is little eccentricity decay.}
  \label{fig:ecc-suite}
\end{figure}

We now determine the radii of the morituri ultimi for eccentric planets. In Figure~\ref{fig:ecc-suite} we show the fates of suites of eccentric planets around a $5\mathrm{\,M}_\odot$ star. The top panel shows the evolution of Jovian planets. These experience significant eccentricity decay prior to engulfment. The bottom shows the evolution of Terrestrial planets. These experience very little eccentricity or semi-major axis decay, and their fate is decided by their small pericentres bringing them within the envelope. The radius of the moriturus ultimus is larger when the effects of eccentricity are included, by up to 0.25 AU for the Jovian planets, up to 0.27 AU for the Neptunians, and up to 0.6 AU for the Terrestrials. In the first case this is attributable to the enhanced semi-major axis decay when eccentricity is included; in the last, to the pericentres' being significantly smaller than the semi-major axis; both factors contribute for the Neptunians. All the critical radii, of both circular and eccentric planets, are shown in Figure~\ref{fig:morituri}.

\subsection{Planets that avoid engulfment}

\begin{figure}
  \includegraphics[width=.5\textwidth]{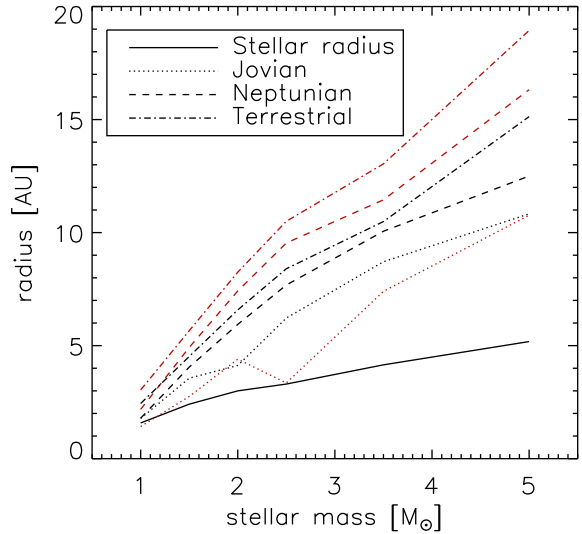}
  \caption{Final radii of the innermost surviving planet, along with the maximum stellar radius on the AGB. Lines are as in Figure~\ref{fig:morituri}.}
  \label{fig:elapsi}
\end{figure}

It is of great interest for predicting the nature of planetary systems around WD stars to determine the final radius of the planet closest to the star that evades engulfment, a planet we call the \emph{elapsus citimus}. These radii are shown in Figure~\ref{fig:elapsi}. They are larger for larger progenitor mass, both because larger stars lose a greater fraction of their mass on the AGB, causing greater adiabatic orbital expansion, and because the larger maximum radii of such stars cause the engulfment of more distant planets. For the Jovian planets, tidal effects are strong enough to strand some planets so that they finish up inside the maximum stellar radius. This effect is larger the smaller the mass of the star, as seen also in \cite{VL09} for the RGB. Such stranding requires a good deal of fine-tuning however, and is infrequent. For this reason, it is also possible that we missed some closer-in planets with our sampling of parameter space (an initial separation of $8\times10^{-4}$\,AU). However, due to the non-smooth nature of the stellar radius evolution, it is not guaranteed that the elapsus citimus can end up arbitrarily close to the star. A planetary eccentricity decreases the radius of the Jovian elapsus citimus: it is pulled in more because of the stronger tidal forces.

\begin{figure}
  \includegraphics[width=0.5\textwidth]{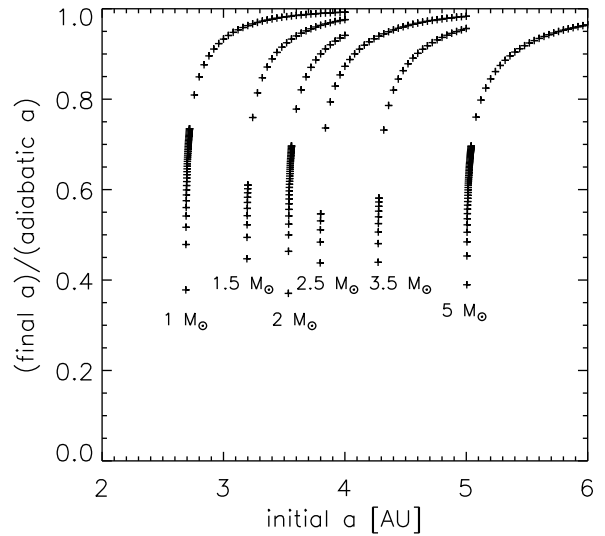}
  \caption{Ratio of planets' final semi-major axis to that expected from adiabatic orbital expansion, as a function of initial semi-major axis. Tidal forces cause the final axes of many planets that are not engulfed to be significantly below those expected from pure adiabatic orbital expansion. Sets of points are in order of increasing stellar mass, from left to right.}
  \label{fig:ratio}
\end{figure}

We show the effects of tidal forces on shrinking the orbit of the elapsus citimus in Figure~\ref{fig:ratio}, which shows the ratio of the final orbit of each surviving simulated planet, of Jovian mass and on a circular orbit, to that expected purely from adiabatic mass loss. Any planet lying within the dotted line of Figure~\ref{fig:morituri} is engulfed by the envelope and not shown. Just outside this radius, tidal orbital decay competes with the orbital expansion due to mass loss, and the orbit of the planet is reduced much beyond that expected from pure adiabatic expansion. For this reason, the radii of the elapsi citimi plotted in Figure~\ref{fig:elapsi} are not simply the radii of the morituri ultimi scaled up by a factor of the fraction of the stellar mass lost. However, for planets with initial orbits beyond around half an AU greater than those of the morituri ultimi, tidal forces are weak and adiabatic orbital expansion is a good approximation.

For the Earth-mass and Neptune-mass planets, tidal effects are weak enough that no stranding inside the maximum stellar radius occurred. A lower planet mass means that the elapsus citimus ends up more distant from the star, although it begins closer than for a more massive planet. This is because the larger planets experience stronger tidal orbital decay. The radius of the Terrestrial elapsus citimus is only just short of the initial radius of the moriturus ultimus scaled up by adiabatic mass-loss. For the low-mass planets, a planetary eccentricity means that the elapsus citimus is at a greater radius than for the circular planets, in contrast to the case for the Jovian planets, because the planetary eccentricity does not add significantly to the weak tidal forces. While post-AGB stars with $1\mathrm{\,M}_\odot$ progenitors will be cleared of planets out to two or three AU, those with $5\mathrm{\,M}_\odot$ progenitors will be cleared of Jovian planets out to 10 AU and of terrestrials out to 15 or more AU.

\subsection{Sensitivity to parameters}

There are four parameters in our tidal model which are poorly known. These are: $c_\mathrm{F}$, governing the onset of the frequency dependence of the tidal response, and $\gamma_\mathrm{F}$ its exponent (Equation~\ref{eq:f}); $\eta_\mathrm{F}$, used to estimate the convective time-scale (Equation~\ref{eq:tconv}); and $f^\prime$ which governs the magnitude of the tidal force (Equation~\ref{eq:f}). For the studies above, we chose $\eta_\mathrm{F}=3$, $c_\mathrm{F}=1$, $f^\prime=9/2$ and $\gamma_\mathrm{F}=2$ for consistency with \cite{VL09}. Now we explore the impact of changing these parameters to other values given in the literature.

\begin{figure}
  \includegraphics[width=.5\textwidth]{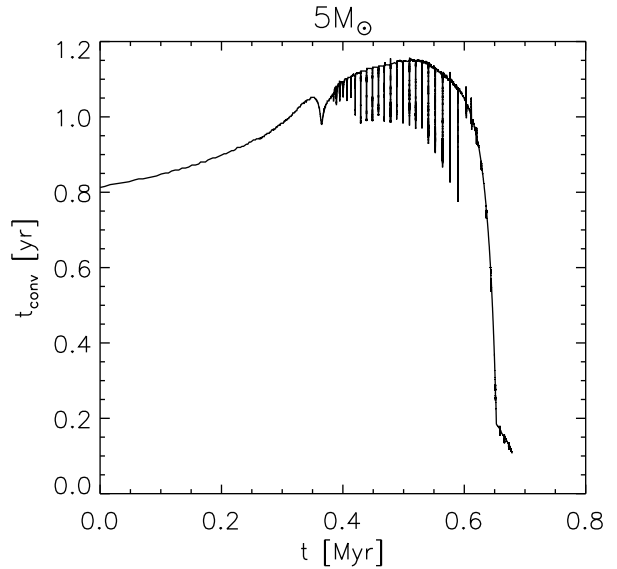}
  \caption{Convective time-scale $t_\mathrm{conv}$ over the thermally pulsing AGB lifetime of a $5\mathrm{\,M}_\odot$ star. The parameter $\eta_\mathrm{F}=3$.}
  \label{fig:tconv}
\end{figure}

The parameters governing the frequency dependence, $c_\mathrm{F}$ and $\gamma_\mathrm{F}$, turn out to be relatively unimportant, because the convective time-scale is usually shorter than the tidal periods (at most $t_\mathrm{conv}$ is 1.15 years, for the $5\mathrm{\,M}_\odot$ star: see Figure~\ref{fig:tconv}). Varying $\gamma_\mathrm{F}$ while keeping the other parameters unchanged thus has no effect on our results. There is a small effect in changing $c_\mathrm{F}$: where we have picked $c_\mathrm{F}=1$, comparing a complete tidal forcing period to the convective time-scale, other suggestions are for $c_\mathrm{F}=2$ \citep{Zahn66} or $c_\mathrm{F}=2\pi$ \citep{GoldreichKeeley77}. Picking the larger as having the strongest effect, we find a decrease in the radius of the moriturus ultimus of around 0.1\,AU for a circular Jupiter and 0.2\,AU for an eccentric Jupiter orbiting a $5\mathrm{\,M}_\odot$ star. The eccentric planet is more strongly affected since the eccentricity brings higher frequency Fourier components into play. This will make an application of the full theory of \cite{EggletonKiselevaHut98} for higher eccentricities impossible, as this theory is inconsistent with the responses to the Fourier components being independent. Dealing with higher eccentricity planets will require explicit high-order eccentricity expansions of the tidal potential.

 There is a similar effect in changing the estimated convective time-scale $t_\mathrm{conv}$ with the parameter $\eta_\mathrm{F}$. Changing $\eta_\mathrm{F}$ from our 3 to an alternative 1 \citep[e.g.,][]{Zahn77} increases the convective time-scale by almost 50\%, weakening tidal forces by increasing the denominators of equations (\ref{eq:adotstar}) and (\ref{eq:edotstar}). In this case, the radius of the Jovian moriturus ultimus decreases by $0.1-0.2$\,AU, depending on stellar mass. We also reduced the multiplicative factor $f^\prime$ by a factor 10, as suggested by the numerical simulations of \cite{Penev+09}, although we caution that these were for Main Sequence Solar-type stars. This resulted in a radical decrease in the radius of the Jovian moriturus ultimus, by 0.67 AU for the $1\mathrm{\,M}_\odot$ star and 1.25 AU for the $5\mathrm{\,M}_\odot$ star.

Finally, we note that the planetary tide had no noticeable effect on the orbital evolution, even when the quality factor was very low: varying $Q^\prime_\mathrm{pl}$ from $10^2$ to $10^{10}$ for giant planets and from $10^0$ to $10^3$ for terrestrials had no effect on the outcome of the integrations. Planetary tides are unimportant due to the extremely small value of $R_\mathrm{pl}/a$.

\section{Discussion}

\label{sec:discussion}

\subsection{Relation to previous work}

Our study naturally follows on from that of \cite{VL09}, studying the fate of planets around RGB stars. \cite{VL09} found that Jovian planets initially within 3 AU of a $1\mathrm{\,M}_\odot$ star would be drawn into the stellar envelope and engulfed. While this is immaterial for determining the maximum initial radius of engulfed planets on the AGB, it may affect the minimum orbital radius after the AGB, since the minimum final radius is attained for some minimum initial radius. If planets at this radius are engulfed on the RGB, the minimum post-AGB radius may be larger. We note, however, that in the integrations of \cite{VL09}, planets were sometimes ``stranded'' by tidal forces, ending up inside the maximum RGB stellar radius due to the small amounts of mass loss and the smooth stellar radius evolution. Furthermore, observations of Horizontal Branch stars have revealed some close companions that have apparently survived the RGB phase \citep{Silvotti+07,Charpinet+11,Setiawan+10}, showing that some planets may be able to survive engulfment in the stellar envelope \citep[e.g.,][]{Soker98,BearSoker12}. Hence, the region closer to the star can be repopulated at the end of the RGB, and this makes our determination of the minimum post-AGB radius sound. For higher-mass stars this is not an issue, since any planets beyond a fraction of an AU are safe on the RGB \citep{VL09}.

We also build naturally on the work of \cite{VL07}, who studied planetary survival around AGB stars but neglected tidal evolution, on the grounds that the maximum stellar radius, and hence the strongest tidal force, exists only briefly at the AGB tip. It is clear from the above results however, that for Jovian planets the tidal forces are indeed significant over much of the AGB phase. \cite{VL07} estimated the radii of the innermost surviving planets by assuming that they would come from planets whose initial radius was at the maximum stellar radius. However, since significant stellar mass loss occurs before the maximum radius is reached, the innermost survivor can originate from inside the maximum radius. Hence, the values for the minimum radii at which planets can be found on leaving the AGB phase in that work, ranging from 3 AU for a $1\mathrm{\,M}_\odot$ star to 30 AU for a $5\mathrm{\,M}_\odot$ star, are somewhat overestimated: We find, on the contrary, 2 and 10 AU respectively. \cite{VL07} however treated another threat to planetary survival: atmospheric evaporation due to the intense XUV radiation during the PN phase. They found that the gaseous envelopes of Jupiter-mass planets ending up within $3-5$ AU will be totally evaporated during the PN phase, and hence that for lower-mass WDs that which determines the survival of Jovian planets is evaporation, not tidal evolution. Note however, that the cores of these planets will remain, and the lower-mass planets we have considered will not be as strongly affected. \cite{VL07} also concluded that for higher-mass stars the survival of planets to the WD phase is not affected by evaporation, due to their very large semi-major axes. Even with our significant reduction of the axis of the innermost survivor around these stars, we can agree that evaporation will not be significant for them.

A study of the combined effects of tides and mass loss over the whole stellar lifetime has recently been carried out by \cite{Nordhaus+10}. This study investigated the fates of Jovian planets, brown dwarfs and low-mass stellar companions to stars of mass $1-3\mathrm{\,M}_\odot$. They compared the tidal formalism of Zahn and the $Q$ formalism, finding that the latter yields considerably weaker tides. This may be the result of their using a $Q$ calibration for Main-Sequence stars, not giants, as we remarked in Section~\ref{sec:eqns}.

\cite{Nordhaus+10} also investigated a range of stellar mass-loss rates, varying the Reimers $\eta$ parameter. However, the commonly-used \cite{Reimers75} rate does not apply
to the AGB phase. As we have
discussed in the previous sections, mass loss is a determining factor in the evolution of the orbit of the planet, and this
is especially true on the AGB where large mass loss takes
place. We therefore spend some time here discussing mass loss on the AGB and its effects on planetary orbital evolution.

\cite{Renzini81} showed that a much stronger wind than the one allowed by the Reimers
parameterization---a ``superwind''---develops at a crucial point in the star's evolution, reaching mass loss
rates of 10$^{-5}$ to 10$^{-4}\mathrm{\,M}_\odot\mathrm{\,yr}^{-1}$. The Reimers relation for the AGB phase
does not agree with observations through the entire AGB phase: it gives mass-loss rates
that increase too slowly, take too long, and reach values too small \citep[see e.g.][]{Willson00,BowenWillson91}. Furthermore, Reimers mass-loss rates do not allow
the formation of planetary nebul{\ae} \citep[e.g.][]{Renzini81,IbenRenzini83,Schoenberner83}, and fail to reproduce the high mass-loss rates observed in AGB stars \citep[e.g.][]{Winters+00,Wood79}, the increase in mass loss during the
AGB ascent \citep{Wood79,Willson00}, and the observed
C-star luminosity functions in the Magellanic Clouds and the
thermally pulsing AGB lifetimes of M and C stars in Magellanic Cloud star clusters \citep{Marigo+08}.

It is important to keep in mind that although mass loss is a crucial process in the evolution of stars along the AGB,
it cannot be calculated from first principles since it requires
the use of dynamical model atmospheres that consider time-dependent dynamics (shock waves and winds), radiation
transfer (strong variable stellar radiation field), and dust and molecular formation processes.
Several stellar evolutionary models that follow the temporal behaviour of the mass loss during the AGB together with
the thermal pulsation of the star are available in the literature \citep{VassiliadisWood93,Blöcker95,Schröder+99,SchröderKuntz05,Wachter+02}. Although the mass-loss rates are not derived
from first principles in these models,
and most of them rely either on the dynamical model atmosphere calculations of \cite{Bowen88} and \cite{Arndt+97},
or on the semi-empirical mass-loss rate formula derivations of \cite{Wood90}, they do provide an opportunity to
study the extensive history of mass loss on the AGB and beyond. The comparison of the different mass-loss
prescriptions with observations of individual stars in the AGB evolutionary phase is complicated given the
variable nature of the star (and the mass loss), and a full discussion on the subject would be lengthy and
beyond the scope of this work. However, the \cite{VassiliadisWood93} prescription
is the most widely used, and not only has it not been ruled out by observations but it has even been favoured over other
parameterizations \citep[see, e.g.,][]{Marigo+98,Ziljstra+02}.

Note also that the maximum mass-loss rate adopted in the \cite{VassiliadisWood93} models is limited to
the radiation pressure limit. Dust-driven wind models allow values to be up to two times this limit. If the maximum
mass-loss rates increase, the timescale of the evolution of the star is modified accordingly. By considering the
evolution of the mass-loss evolution provided by a set of stellar evolution calculations, we are simulating a realistic
scenario with no free parameters that allows a more self-consistent treatment of the problem.

\begin{figure}
  \includegraphics[width=.5\textwidth]{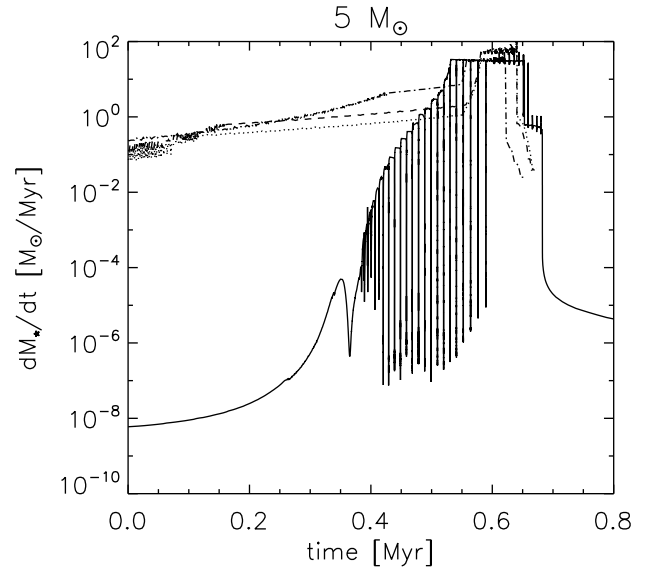}
  \caption{Mass-loss rates for $5\mathrm{\,M}_\odot$ stars under our model (solid) and SSE models with $\eta=0.5$ (dotted), $1.0$ (dashed) and $5.0$ (dot-dashed).}
  \label{fig:massloss}
\end{figure}

To compare our mass-loss prescription with the Reimers prescription, we show the mass-loss rate from our $5\mathrm{\,M}_\odot$ model in Figure~\ref{fig:massloss}. We also show on this plot the mass-loss rates from models using the SSE code \citep*{HurleyPolsTout00} with Reimers mass-loss rates with parameter $\eta=0.5,$ $1.0$ and $5.0$. Under the Reimers prescription, more mass loss occurs earlier on the AGB than in our chosen models regardless of the chosen value of $\eta$ in the SSE models. Since mass loss occurs earlier using the Reimers prescription, planets can move out of reach of tidal forces before these have a chance to act.

In the context of comparing our results to those of \cite{Nordhaus+10}, we now note the following points. First, the \cite{Nordhaus+10} stellar models fail to reproduce the initial-to-final mass relation using the Reimers prescription: note that  their $3\mathrm{\,M}\odot$ star ends the AGB with a mass of $1.1$ or $1.0\mathrm{\,M}_\odot$ with $\eta = 0.7$ and 1 (J. Nordhaus, 2012, private communication), with only $\eta=5$ approaching the canonical value of $0.7\mathrm{\,M}_\odot$ \citep[e.g.,][]{Weidemann87,Kalirai+08}. One should then expect that, using a comparable formalism for the tides, our morituri ultimi should agree better with the \cite{Nordhaus+10} determination of maximum semi-major axis that is tidally engulfed which uses the largest (and unrealistic) value of the mass-loss  parameter $\eta = 5$. We find however the opposite; that is, our results using the Zahn formalism give comparable values for the radii of the  morituri ultimi only for the \cite{Nordhaus+10} calculations with Reimers $\eta = 0.7$ and 1, i.e., those that give an unrealistic final mass. Second, our results using the Zahn formalism give somewhat larger values for the radii of the morituri ultimi for lower-mass stars (for Jovian planets around $1\mathrm{\,M}_\odot$ stars, 2.7\,AU versus 1.6\,AU) but similar values for higher-mass stars (for Jovian planets around $2.5\mathrm{\,M}_\odot$ stars, 3.8\,AU versus $\sim 4$\,AU). We believe that the reasons behind this large discrepancy are as follows. First, under the Reimers prescription, mass loss is continuous from early times, and hence planets' orbits can expand before tides have a chance to act, as described above. Second, the \cite{Nordhaus+10} models have a higher final stellar mass and a smaller maximum radius than ours. Both of these mean that tidal forces are weaker in the \cite{Nordhaus+10} study than in ours, and hence that the radius of the moriturus ultimus is lower. For higher-mass stars, however, the final stellar mass in the \cite{Nordhaus+10} models is much larger than it ought to be, and therefore the orbital expansion due to mass loss is less. Hence both the tidal decay and the adiabatic orbital expansion are weakened, which conspire to give a radius of the moriturus ultimus similar to ours.

The stellar evolutionary model is one of two major sources of uncertainty in our calculations, the other being the tidal model. It may be asked which is the more important. For $1\mathrm{\,M}_\odot$ stars, our moriturus ultimus radius of $2.7$\,AU is considerably larger than those of \cite{Nordhaus+10} using the same tidal prescription ($0.6$ to $1.9$\,AU, $1.6$\,AU for the same final White Dwarf mass). The biggest single-parameter change in the tidal model---changing $f^\prime$ to align with the hydrodynamical simulations of \cite{Penev+09}---gives a change of $0.7$\,AU for this stellar mass. Hence, it appears that, despite uncertainties in tidal theory, a significant improvement can indeed be obtained by improving the stellar models.

\subsection{Implications}

We have shown that, around WDs whose progenitor masses range from 1 to $5\mathrm{\,M}_\odot$, planetary companions may be expected outwards of 1.4 AU to 10 AU for Jovian planets around $1\mathrm{\,M}_\odot$ stars to $5\mathrm{\,M}_\odot$ stars, and outwards of $2.5$ AU to $15$ AU for Terrestrial planets around $1\mathrm{\,M}_\odot$ stars to $5\mathrm{\,M}_\odot$ stars. The closer planets may experience significant evaporation, yet their cores may remain. If engulfed in the stellar envelope, low-mass planets should not have much chance of survival. However, planets of higher mass than those considered here, or their remains, might survive the Common Envelope phase and be found interior to the limits provided in this work \citep[e.g.,][]{BearSoker12}. The distant planets that we expect to survive may be detectable by direct imaging, but only if massive enough \citep{Burleigh+02}. Timing methods \citep[e.g.,][]{Silvotti+11} and transits \citep[e.g.,][]{Agol11,Faedi+11} are more sensitive to close-in planets, for which survival of the AGB phase may be a challenge. However, we do not wish to discourage observers from trying to find planets within these limits given that other mechanisms than those considered here might be at work and that successful discoveries will only stimulate further theoretical work.

Any planets that do survive the AGB phase can still have a significant dynamical influence on any other bodies such as remnant comets or asteroids in the system, and in this way can bring about the delivery of pollutant elements to the neighbourhood of the WD. Such delivery mechanisms require dynamical instability, whether local instability of the metalliferous planetesimals \citep{BMW11,BW12,DWS12} or wholesale global instability \citep[][D.~Veras et al., MNRAS, submitted]{DS02}.

It may be thought that the stability of the system that survives the AGB is enhanced by the effects of tidal forces destroying the inner planet, or separating it from those further out. Multiple-planet scattering, as studied by \cite{DS02}, acts on a time-scale strongly determined by the separation of the planets \citep{Chambers96,FQ07}, and the detachment or destruction of an inner planet would stabilize a system liable to undergo scattering. And a planet scattering particles inwards from an exterior belt, as envisaged by \cite{BMW11}, can be drawn away from the planetesimals by the tides.

\begin{figure}
  \includegraphics[width=.5\textwidth]{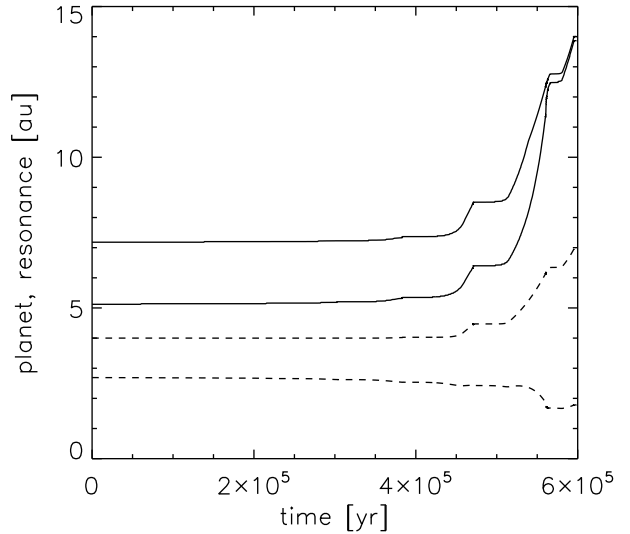}
  \caption{Change in location of Jovian planets (dashed lines) and secular resonances (solid lines) around a $1\mathrm{\,M}_\odot$ star.}
  \label{fig:secres}
\end{figure}

However, in other cases the detachment of an inner planet may make a system more prone to instability. Multiple-planet systems with inclined binary companions can be stabilized by their mutual Laplace--Lagrange perturbations against the otherwise devastating Kozai effect \citep{Innanen+97,SalehRasio09}, a protective mechanism possibly at work in 55~Cnc \citep{Kaib+11}. If an inner planet is detached, the coupling between the planets may weaken enough to trigger instability after the AGB phase. In any system, the detachment of an inner planet may cause divergent crossing of Mean Motion Resonances and the sweeping of secular resonances, both of which can pump up the eccentricities of planetesimals, and which are thought to have had an important effect on bodies in the early Solar System \citep[e.g.,][]{Tsiganis+05,Brasser+09}. Finally, the closer the inner planet to the WD, the easier can be the delivery of small bodies being scattered by chains of planets \citep{BW12}.

 As an example of the movement of secular resonances, we can consider a system with two planets of Jovian mass, one initially at 2.69\,AU and one at 4\,AU, around a $1\mathrm{\,M}_\odot$ star. Their final semi-major axes are 1.78\,AU and 6.96\,AU. As the system evolves, the secular resonances move from 5.1\,AU and 7.2\,AU to 13.9\,AU and 14.0\,AU (see Figure~\ref{fig:secres}). A final semi-major axis of 14\,AU corresponds to an initial semi-major axis of 8\,AU, and so the secular resonances move out faster than orbits expand due to adiabatic mass loss. Hence, any planetesimals initially between 5 and 8\,AU will be swept by one or both resonances, pumping up their eccentricity and potentially rendering them vulnerable to scattering at the end of the AGB itself or beyond. The full working out of the system-wide effects of the tidal evolution of inner planets must, however, be deferred for future work.

\section{Conclusions}

\label{sec:conclusions}

We have studied the orbital evolution of planets around Asymptotic Giant Branch stars ranging from $1\mathrm{\,M}_\odot$ to $5\mathrm{\,M}_\odot$ as the star expands and loses mass, with detailed stellar models giving the stellar mass loss and radius evolution. We considered planets of Jovian mass and lower, which have insufficient energy to unbind the stellar envelope and will likely be destroyed upon entering it. To compute the tidal forces we used the formalism of \cite{Zahn77}, appropriate for AGB stars with massive convective envelopes. Planetary orbital evolution is then a contest between tidal forces pulling the planet in and mass loss causing orbital expansion.

Jovian planets experience strong tidal forces, and Jovian planets that begin on orbits outside the maximum stellar radius can be drawn into the envelope. The maximum initial radius at which a Jovian planet on a circular orbit can be drawn into the envelope ranges from 2.6\,AU for a $1\mathrm{\,M}_\odot$ star to 5\,AU for a $5\mathrm{\,M}_\odot$ star. Lower mass planets feel tidal forces less strongly, and planets that begin on orbits interior to the maximum stellar radius can escape engulfment due to their orbits expanding under stellar mass loss: Terrestrial planets on initially circular orbits can be drawn into the envelope at initial radii of up to 1.5\,AU to 2.8\,AU over our mass range.

The inclusion of a planetary eccentricity increases the maximum initial radius for which planets can enter the stellar envelope. In the case of Jovian planets this is due to the strengthened tidal forces: the lifetime of an eccentric planet is reduced by about 10--20\% relative to a circular one, and eccentricity decays almost to zero as the planet plunges into the envelope. Eccentricity is damped entirely by the stellar tide and planetary tides are ineffective. The eccentricities of Terrestrial planets barely decay, and their having smaller pericentres relative to circular planets means that the maximum initial radius at which planets will be engulfed is much larger, and their lifetime much shorter, by as much as one half.

We then found the minimum semi-major axis at which planets may be found around White Dwarfs after the AGB stage has ended. This is larger for more massive stars, both because larger stars lose a greater fraction of their mass on the AGB, causing more adiabatic orbital expansion, and because for the larger stars the initial radii of the survivors are larger. Assuming that any planet of Jovian mass or less entering the stellar envelope is destroyed, we do not expect to find Terrestrial planets within 2\,AU of WDs of progenitor mass $1\mathrm{\,M}_\odot$ or within 15\,AU of WDs of progenitor mass $5\mathrm{\,M}_\odot$. Jovian planets, which can have their adiabatic orbital expansion retarded by tidal forces even as they avoid the stellar envelope, may be found somewhat closer, at 1.5\,AU or 10\,AU respectively. However, a certain fine-tuning is required in order that such planets avoid plunging into the envelope while still experiencing a non-negligible tidal force: there is a region around 0.5\,AU in width outside the orbit of the maximum orbit that gets engulfed in which Jovian planets still experience strong tidal orbital decay.

We studied the effects of changing the several parameters in the Zahn tidal theory. The systems we studied have convective time-scales a few times shorter than orbital time-scales, and so the unknown nature of the frequency dependence of the tidal forces are relatively unimportant. However, uncertainties in the overall magnitude of the tidal response can have a significant effect, with a reduction of a factor 10 to align the Zahn theory with the numerical simulations of \cite{Penev+09} reducing the radius at which Jovian planets can be engulfed by around 1\,AU.

Finally, we noted that in multi-planet systems the differential orbital evolution of more and less distant planets could have significant implications for the stability of the system on the White Dwarf phase, and therefore for the pollution of the White Dwarf's atmosphere and its environment by destroyed planets or planetesimals.

\acknowledgements

This work is funded by the Spanish National Plan of R\&D grant AYA2010-20630, ``Planets and stellar evolution''. EV also acknowledges the support provided by the Marie Curie grant FP7-People-RG268111. AJM would like to thank Adrian Barker for useful discussions on tidal theory. We should like to thank Dimitri Veras and the anonymous referee for suggesting improvements to the manuscript.

\bibliographystyle{apj}

\end{document}